\documentclass[12pt,a4paper]{article}
\usepackage{a4wide}
\usepackage[centertags]{amsmath}
\usepackage{amssymb}
\usepackage{epsfig}
\usepackage{psfig}
\usepackage{cite}
\begin{document}
\thispagestyle{empty}
\begin{flushright}
IPPP/05/78\\
DCPT/05/156\\
\end{flushright}
\vskip 2cm

\begin{center}
{\huge The Higgs Mechanism in Heterotic Orbifolds}
\end{center}
\vspace*{5mm} \noindent
\vskip 0.5cm
\centerline{\bf Stefan F\"orste$^a$, Hans Peter Nilles$^{b}$ and Ak\i{}n
  Wingerter$^c$}
\vskip 1cm
\centerline{$^a$
\em Institute for Particle Physics Phenomenology (IPPP)}
\centerline{\em South Road, Durham DH1 3LE, United Kingdom}
\vskip 0.5cm
\centerline{$^b$\em Physikalisches Institut, Universit\"at Bonn}
\centerline{\em Nussallee 12, D-53115 Bonn, Germany}
\vskip 0.5cm
\centerline{$^c$ \em Department of Physics, The Ohio State University}
\centerline{\em 191 W.\ Woodruff Ave., Columbus, OH 43210, USA}
\vskip2cm

\centerline{\bf Abstract}
\vskip .3cm

We study spontaneous gauge symmetry breaking in the framework of orbifold
compactifcations of heterotic string theory. 
In particular we investigate the electroweak symmetry breakdown via
the Higgs mechanism. 
Such a breakdown can be achieved by continuous Wilson lines. 
Exploiting the geometrical properties of this scheme we develop a new
technique which simplifies the analysis used in previous discussions. 

\vskip .3cm

\newpage

\section{Introduction}

An outstanding problem of string theory is the construction of
realistic models. One of the first attempts to solve that problem is
provided by orbifold compactifications of the heterotic string
\cite{Dixon:1985jw,Dixon:1986jc}. Such constructions allow us to obtain
four dimensional effective theories with $N=1$ supersymmetry. The
massless spectrum can be determined explicitly and naturally comes out
to be chiral. Further progress has been made by the introduction of
discrete Wilson lines \cite{Ibanez:1986tp}. That mechanism is a
suitable tool in order to control the number of families as well as to
reduce the unbroken gauge group down to groups like
SU(3)$\times$SU(2)$\times$U(1) \cite{Ibanez:1987sn}. Subsequently many
models have been studied \cite{Bailin:1986pd, Bailin:1987pf, 
  Bailin:1987xm, Bailin:1987dm, Ibanez:1987pj, Casas:1987us, Casas:1988se,
  Casas:1988hb, Katsuki:1988ku,
  Katsuki:1989kd, Katsuki:1989ra, 
  Hwang:2002hg, Kim:2003ch, Choi:2003pq, Choi:2003ag, Kim:2003hr,
  Kim:2004pe, Giedt:2001zw, Giedt:2003an,
  Giedt:2005vx,Choi:2004wn}. For a review see e.g.\
\cite{Quevedo:1996sv}. (Related free fermionic constructions
\cite{Faraggi:1992fa, Faraggi:1991jr, Faraggi:2004rq, Donagi:2004ht,
  Faraggi:2004xf} 
will not be considered in the present paper.)  

Bottom up approaches to physics beyond the Standard Model started to
discuss the existence of extra dimensions seriously during the past
years. One of the motivations is that models with extra dimensions
allow to keep attractive features of grand unification while some
severe problems such as the doublet triplet splitting problem can be
eliminated. Models of that type are known as orbifold GUTs
\cite{Kawamura:2000ev, Kawamura:2000ir, Altarelli:2001qj, Kawamoto:2001wm,
  Hebecker:2001wq, Asaka:2001eh}. See \cite{Hall:2002ea} for a review
containing more references.

The observations within bottom up approaches triggered renewed
interest in heterotic orbifold constructions.
Indeed, combining the top down string theory constructions with the
bottom up orbifold GUT models has the prospect of providing very
useful results. Field theory constructions can be put on firmer grounds since
string theory yields consistent prescriptions how to deal with orbifold
geometries and serves as a UV completion. Guidelines in GUT orbifold
model building can emerge from string theory. On the other hand, 
results of bottom up approaches improve the situation for attempts to
connect string theory to the real world. From all possible candidates
heterotic orbifolds are the most natural ones incorporating an
effective orbifold GUT picture. Moreover, such a 
picture is obtained in models where the orbifold possesses fixed
tori. Together with the requirement of unbroken $N=1$ supersymmetry
this restricts the orbifold group to be either ${\mathbb Z}_M \times
{\mathbb Z}_N$ or ${\mathbb Z}_K$ with $K$ not prime. Recent
investigations focus on ${\mathbb Z}_2 \times {\mathbb Z}_2$
\cite{Forste:2004ie} and ${\mathbb Z}_6$-II \cite{Kobayashi:2004ud,
  Kobayashi:2004ya,  Buchmuller:2004hv, Buchmuller:2005jr}  models.  

The present paper is devoted to symmetry breaking by continuous Wilson
lines within that class of models.
In \cite{Ibanez:1987xa} and more recently in \cite{Forste:2005rs} it
was demonstrated that continuous
Wilson lines are most easily included if the orbifold group is
embedded into the gauge group as a rotation of the root lattice. In
this case the Wilson line can be taken to point along Cartan
directions. It was, however, also encountered that it can be
complicated to identify a consistent algebra automorphism which is
induced by the rotational embedding of the orbifold group.

In section 2 we explore to what extend the
rotational embedding of the orbifold group into the gauge group is
essential. We will find that in the field theory picture it is
possible to describe continuous Wilson lines also when the orbifold
group is embedded with the usual shift leaving all Cartan generators
invariant and multiplying root generators with a phase. Thus the shift
embedding refers to a particular choice of the Cartan-Weyl basis.

We will confirm the geometric picture described in
\cite{Forste:2005rs} from a
slightly different angle. At all fixed points the orbifold is defined
by a shift embedding. However, the choice of the Cartan-Weyl basis can
differ from fixed point to fixed point. In particular the embedding of
the Cartan torus into the gauge group can be different. In such a case
the rank of the unbroken gauge group will be reduced.

In section 3 we will rederive the particular symmetry breakings
discussed in \cite{Forste:2005rs} using the new technique. This will
allow us to present the projection patterns for bulk matter in the
considered model.

The rest of the paper is devoted to electroweak symmetry breaking via
continuous Wilson lines. After discussing our general strategy in
section 4 we will consider a ${\mathbb Z}_6$-II model in section 5 and
a ${\mathbb Z}_2 \times {\mathbb Z}_2$ model in section
6. A summary and implications of our findings for string theoretical
model building will be presented in a concluding section 7. A couple of
appendices provides technical details and supplementary informations.   

\section{Wilson Lines in Shift Embeddings}\label{sec:wl}

For simplicity we will discuss a ${\mathbb Z}_2$ orbifold GUT with one extra
dimension in this section. The ${\mathbb Z}_3$ example with two extra
dimensions is presented in appendix \ref{ap:z3}. As far as the computation of
the massless spectrum is concerned a lift to string theory should be
also straightforward. As will become clear shortly one can employ the
method of constructing fixed point equivalent models
\cite{Gmeiner:2002es} which are all
shift embedded without Wilson lines.

The five dimensional theory is a gauge
theory with gauge group $G$.  The coordinate of the extra dimension
is called $x^5$. An effective four dimensional theory arises upon
compactification on an orbicircle $S^1/{\mathbb Z}_2$. This geometry
is obtained by identifying points on the real line which are mapped
onto each other under an element of the space group. The space group
is generated by a ${\mathbb Z}_2$
\begin{equation}
\left( \theta , 0\right):\,\,\, x^5 \to -x^5
\end{equation}
and the $S^1$ compactification
\begin{equation}
(1,e):\,\,\,  x^5 \to x^5 + 2 \pi R .
\end{equation}
Fixed points under the ${\mathbb Z}_2$ action are points which differ
from their ${\mathbb Z}_2$ image by an integer multiple of $2 \pi R$,
i.e.\ they are identical on $S^1$. On $S^1$ there are two fixed points
given by 
\begin{equation}
x^5 = 0 \,\,\, \mbox{and}\,\,\, x^5 = \pi R .
\end{equation} 
We specify the gauge group $G$ by choosing a Cartan-Weyl basis, i.e.\
a set of Cartan generators
\begin{equation}
H_i, \,\,\, i = 1, \ldots , r 
\end{equation}
and a set of root generators
\begin{equation}
E_{\alpha_k}, \,\,\, k = 1, \ldots , \mbox{dim}(G) - r ,
\end{equation}
where $r$ denotes the rank of the gauge group.

The shift embedding of the ${\mathbb Z}_2$ is defined as follows
\begin{equation}
\left( \theta , 0\right):\,\,\, \begin{array}{l l l l}
H_i & \to & H_i , & i = 1, \ldots, r \\
E_{\alpha_k} & \to & \mbox{exp}\{ 2 \pi i \alpha_k \cdot V\}
E_{\alpha_k}, & k 
  = 1, \ldots , \mbox{dim}(G) - r \end{array} ,
\end{equation}
where $V$ is an $r$ dimensional Vector with components $V^i$. The
action on the root 
generators can be alternatively described as
\begin{equation}
E_{\alpha} \to e^{2\pi i V^i H_i} E_\alpha
e^{-2 \pi i V^i H_i} . 
\end{equation}
The requirement that
$\left(\theta^2 ,0\right) = (1, 0)$ leaves $G$ invariant yields the
condition that $2V$ should be an element of the coroot lattice of
$G$. 

The effect of a Wilson line is that the embedding\footnote{Here, we
  have to specify that we first apply $x^5 
  \to - x^5$ and afterwards shift by $2 \pi R$.} of $\left( \theta
,e\right)$ contains an
additional constant gauge transformation $t \in G$
\begin{equation} \label{eq:wlgen}
\left( \theta , e\right):\,\,\, \begin{array}{l l l l}
H_i & \to & t H_i t^{-1} , & i = 1, \ldots, r \\
E_{\alpha_k} & \to & \mbox{exp}\{ 2 \pi i \alpha_k \cdot V\}
t E_{\alpha_k}\, t^{-1}, & k 
  = 1, \ldots , \mbox{dim}(G) - r \end{array} ,
\end{equation}
Applying this transformation twice yields for example
\begin{equation}
\left( \theta , e\right) ^2: \,\,\, H_i \to \tilde{t} t H_i
\,\, t^{-1} \, \tilde{t}^{-1}, 
\end{equation}
where $\tilde{t}$ is the ${\mathbb Z}_2$ image of $t$
\begin{equation}
\tilde{t} =  e^{2\pi i V^i H_i}\, t\, e^{-2 \pi i V^i H_i} .
\end{equation}
Consistency requires\footnote{A further constraint can be derived by
  taking the hermitian conjugate of (\ref{eq:wlgen}) and requiring
  that the relations $H_i = H_i ^\dagger$ and $E_\alpha ^\dagger =
  E_{-\alpha}$ are respected. This gives the condition that $t$ should
  be unitary.} 
\begin{equation}\label{eq:cons}
\tilde{t} t = \pm 1.
\end{equation}
The lower sign appears because gauge fields transform in the adjoint
representation of $G$. Since later we may want to add matter
transforming in different representations we will ignore this
possibility in the following. First, let us discuss the known case
where $t$ is in the Cartan subgroup. Then, (\ref{eq:cons}) leads to a
quantisation condition
\begin{equation}
t \in \mbox{CSG} \Longrightarrow t^2 =1 .
\end{equation}
Writing $t$ as
\begin{equation}\label{eq:gendis}
t = e^{2\pi i T} \,\,\, , \,\,\, T = a^i H_i .
\end{equation}
the condition $t^2 = 1$ implies that twice the vector $a$ (with
components $a^i$) should be an element of the coroot lattice of
$G$. Thus this solution corresponds to the well known case of a {\it
  discrete} Wilson line \cite{Ibanez:1986tp}.

The next solution we want to discuss is
\begin{equation} \label{eq:contsol}
\tilde{t} = t^{-1}.
\end{equation}
This implies $\tilde{T} = -T$ and  corresponds to a {\it continuous}
Wilson line. Defining a conjugated Cartan-Weyl basis
\begin{eqnarray}
\hat{H}_i &  = & t^{1/2} H_i \, t^{-1/2}, \\
\hat{E}_{\alpha} & = & t^{1/2}\, E_\alpha t^{-1/2}
\end{eqnarray}
it is easy to see that (\ref{eq:wlgen}) can be rewritten as
\begin{equation} \label{eq:rotshift}
\left( \theta , e\right):\,\,\, \begin{array}{l l l l}
\hat{H}_i & \to & \hat{H}_i , & i = 1, \ldots, r \\
\hat{E}_{\alpha_k} & \to & \mbox{exp}\{ 2 \pi \alpha_k \cdot V \}
\hat{E}_{\alpha_k}, & k 
  = 1, \ldots , \mbox{dim}(G) - r \end{array} .
\end{equation}
Since these relations will be important in the following let us derive them
in detail. The first line of
(\ref{eq:rotshift}) is obtained as follows. First we observe that   
\begin{equation}
\left(\theta , e\right):\,\,\, \hat{H}_i = t^{1/2} H_i\, t^{-1/2} \to
\tilde{t}^{1/2}\, t H_i \, t^{-1}\, \tilde{t}^{-1/2} .
\end{equation} 
Using (\ref{eq:contsol}) to express $\tilde{t}$ in terms of $t$ gives
the desired result. The second line in (\ref{eq:rotshift}) follows in
a completely analogous way. Note also that the conjugation with
$t^{1/2}$ takes a Cartan-Weyl basis to another Cartan-Weyl basis
implying in particular that an alternative expression for the second
line in (\ref{eq:rotshift}) is
\begin{equation}
\hat{E}_{\alpha} \to e^{2\pi i V^i \hat{H}_i} \hat{E}_\alpha
e^{-2 \pi i V^i \hat{H}_i} . 
\end{equation}

We can now describe the geometric picture of Wilson line symmetry
breaking. In all cases the projections at the fixed points are given
as shift embeddings. A discrete Wilson line changes the shift
embedding by replacing $V$ with $V+a$. 
For the continuous Wilson line the shift embeddings at the two fixed
points are both given by the same shift vector $V$. The difference is
in the Cartan-Weyl basis the shift embedding refers to, namely $H$ and
$\hat{H}$, respectively. The embedding
of the Cartan-Weyl basis at the fixed point $x^5 = \pi R$ is
continuously rotated by the Wilson line. In particular the embedding of
the Cartan torus is rotated ($t$ cannot be generated by the $H_i$ in
the case of a continuous Wilson line). The unbroken gauge group in
four dimensions is generated by elements of the Lie algebra of $G$
which are invariant under the projections at all fixed points. For a
continuous Wilson line the rank of the gauge group is reduced. 

To close this section let us give a prescription on how to
parameterise the general solution to (\ref{eq:contsol}).
In this context, the term `general' deserves some discussion. The
Wilson line $T$ can be also viewed as a vacuum expectation value for
the internal gauge field component $A_5$ \cite{Ibanez:1986tp}. We do
not want to count gauge equivalent vacua. In order to be a solution to
(\ref{eq:contsol}) $T$ has to be a linear combination of generators
which are not invariant under the orbifold group. All gauge
inequivalent solutions can be found by identifying a set of
hermitian and mutually commuting operators $C_1, \ldots , C_n$ within the
non-invariant generators. The continuous Wilson line is then a linear
combination of those operators
\begin{equation} \label{eq:wilsol}
T = \sum_{i=1}^n\lambda_i C_i \,\,\, , \,\,\, \mbox{$\lambda_i$ real}.
\end{equation}
One can justify this prescription by establishing the connection to
the case that the orbifold is embedded as a rotation. Within the bulk
gauge algebra one can identify a Cartan subalgebra containing $C_1,
\ldots , C_n$. With respect to that Cartan subalgebra the orbifold
acts as a rotation. $C_1, \ldots , C_n$ are non-invariant Cartan
operators.  Eq.\ (\ref{eq:wilsol}) corresponds to the
prescription known from rotational embeddings
\cite{Ibanez:1987xa,Forste:2005rs}. (The above argument may also be
used for a systematic construction of rotational embeddings.)
 
For completeness, let us mention also a solution to (\ref{eq:cons}) (with
the upper sign) corresponding to a superposition of a continuous and a
discrete Wilson line. This is given by
\begin{equation}
t= e^{2\pi\,i\,\left( T_1 + T_2\right)} ,
\end{equation}
with
\begin{equation}
T_1 = a^i H_i , 
\end{equation}
where $a$ is an element of the coroot lattice (as in
(\ref{eq:gendis})) and
\begin{equation}
\tilde{T}_2 = - T_2 .
\end{equation}
So far, we have superposed our solutions for a discrete and a
continuous Wilson line. In order, that this superposition solves
(\ref{eq:cons}) we need to satisfy that the discrete and continuous
Wilson line commute
\begin{equation}
\left[ T_1 , T_2\right] = 0.
\end{equation}

\section{Continuous Wilson Line Breaking}

The scales for symmetry breaking by the gauge embedding of orbifold twist
and discrete Wilson lines 
is given by the compactification scale. The continuous Wilson line
corresponds to a VEV along a flat direction
\cite{Font:1988tp,Font:1988mm} and 
hence the breaking scale can be adjusted continuously between zero and the
compactification scale. In realistic scenarios the flat direction
should be lifted e.g.\ by SUSY breaking. This goes beyond our present
investigations but we anticipate that depending on the SUSY breaking
mechanism there is some choice for the breaking scale due to
continuous Wilson lines. In the present section we consider the
possibility that this is a scale where a Pati-Salam
gauge symmetry is broken to the Standard Model gauge group.
In particular we will rederive the model discussed in
\cite{Forste:2005rs} using the technique introduced in the previous
section. In \cite{Forste:2005rs}  a six dimensional theory
with gauge group $G= \mbox{E}_6$ was compactified on a $T^2/{\mathbb Z}_2$
orbifold.  The E$_6$ was broken by the
orbifold to SO(10)$\times$U(1). A continuous Wilson line along the sixth
direction was used to break this to SU(5)$\times$U(1). Other possible
breakings were given as SO(10)$\times$U(1) to SO(7)$\times$U(1) or
SU(4)$\times$U(1). A second discrete Wilson line along the fifth
direction resulted in a further symmetry breaking (see figure
\ref{fig:6dsetup} for 
a reminder). We will come back
to the discrete Wilson line later and first check whether the
alternative technique given 
in the previous section yields the same SO(10)$\times$U(1)
breaking patterns as in \cite{Forste:2005rs}, for the continuous
Wilson line. 

\begin{figure}[h!]
\centering
\begin{center}\input 6dsetup.pstex_t \end{center}
\caption{The setup of \cite{Forste:2005rs}.}
\label{fig:6dsetup}
\end{figure}
\subsection{Gauge Symmetry Breaking}
\subsubsection{Breaking E$_6$ to SO(10)$\times$U(1)}

As in \cite{Forste:2005rs} we embed the E$_6$ root vectors into an
eight dimensional space by writing the six Cartan generators into the
following eight dimensional vector
\begin{equation} \label{eq:eightcart}
H = \left( H_1 , H_1 , H_1, H_2 , H_3 , H_4 , H_5 , H_6 \right) .
\end{equation}
The E$_6$ roots are also written as eight dimensional vectors where
each entry gives the eigenvalue of the root operator under the adjoint
action of the Cartan 
generator appearing at the same position in (\ref{eq:eightcart}). The
72 root vectors of E$_6$ are then
\begin{itemize}
 \item 40 root vectors of the form
$$ \left( 0,0,0, \underline{\pm 1, \pm 1, 0,0,0}\right) ,$$
where underlined entries can be permuted,
\item 32 root vectors of the form
$$ \left( \pm \left(\frac{1}{2}, \frac{1}{2}, \frac{1}{2}\right),
  \pm \frac{1}{2}, \pm \frac{1}{2}, \pm \frac{1}{2}, \pm \frac{1}{2},
  \pm \frac{1}{2} \right) $$
with an even number of minus signs.
\end{itemize}
In order to break E$_6$ to SO(10)$\times$U(1) we use the shift vector
\cite{Kobayashi:2004ya} 
\begin{equation}\label{eq:rabyshift}
V = \left( 1 ,1, 1, 0,0,0,0,0\right) .
\end{equation}
The spinorial root vectors listed in the second item above have half
integer scalar product with $V$. The corresponding root operators are
projected out at the fixed point $x^6 = 0$. This yields the unbroken
gauge group SO(10)$\times$U(1) where the U(1) factor is generated by
$H_1$. 

\subsubsection{Continuous Wilson Line Breaking}\label{sec:wlbreaking}

As discussed in section \ref{sec:wl} the effect of the continuous
Wilson line is given by conjugating the Cartan-Weyl basis with
$t^{1/2}$ and 
taking the same shift embedding with respect to the conjugated
basis. This will give also SO(10)$\times$U(1) unbroken gauge symmetry
at the fixed point $x^6 = \pi R_6$. In order to find the unbroken
gauge group in four dimensions we have to specify $t$ and eliminate
those group elements which do not commute with $t$. Our definition for
the continuous Wilson line was that 
\begin{equation}
t = e^{2\pi i T}
\end{equation}
is mapped onto its inverse under the shift embedding, or in other
words that $T$ is a superposition of root generators corresponding to
spinorial roots. In order to find a suitable parameterisation we
proceed as discussed in the end of section \ref{sec:wl}. A maximal set 
of hermitian and commuting operators is given by\footnote{ Some
  tentative thoughts about such operators in the context of rank
  reduction can be found in \cite{Hebecker:2003jt}.}   
\begin{equation}
C_1 = E_\delta + E_{-\delta}\,\,\, , \,\,\, C_2 = E_\gamma +
E_{-\gamma},
\end{equation} 
with
\begin{eqnarray}
\delta & = &\left(\frac{1}{2},\frac{1}{2},\frac{1}{2},\frac{1}{2},
\frac{1}{2},\frac{1}{2},\frac{1}{2},\frac{1}{2}\right)
\label{eq:delta} \\ 
\gamma & = & \left(-\frac{1}{2},-\frac{1}{2},-\frac{1}{2},-\frac{1}{2},
\frac{1}{2},\frac{1}{2},\frac{1}{2},\frac{1}{2}\right) .   
\end{eqnarray}
Then the general continuous Wilson line can be written as
\begin{equation}\label{eq:genconwile6}
T= \lambda_1 \left(E_\delta + E_{-\delta}\right) + \lambda_2 \left(
E_\gamma + E_{-\gamma}\right) .
\end{equation}
For completeness let us also give a set of E$_6$ Cartan generators
with respect to which the orbifold is embedded as a rotation:
\begin{equation}
H_1 - H_2 ,\,\,\, H_3 - H_4, \,\,\, H_3 - H_5, \,\,\, H_3 -H_6, \,\,\,
E_\delta + E_{-\delta}, \,\,\, E_{\gamma} + E_{-\gamma} .
\end{equation}
Now we will discuss the four dimensional unbroken gauge
symmetries. 

\vskip 1cm 

\centerline{\it Generic Case: $\lambda_1 \not= 0$, $\lambda_2 \not=
  0$, $\lambda_1 \not= \pm\lambda_2$}

\vskip 0.5cm

First, we take $\lambda_1$ and $\lambda_2$ to be
independent and non vanishing. In this case the following SO(10)
generators are invariant under conjugation with $t$:
\begin{itemize}
\item four Cartan generators: $H_1 - H_2$, $H_3 - H_4$, $H_3 - H_5$,
  $H_3 - H_6$,
\item 12 root generators corresponding to root vectors of the form
$\pm \left( 0,0,0,0,\underline{1, -1, 0,0}\right)$.  
\end{itemize}
These generate an SU(4)$\times$U(1) symmetry. The rank
of the gauge group has been reduced by two.   

\vskip 1cm

\centerline{\it Special Case: $\lambda_1 \not= 0$, $\lambda_2 \not=
  0$, $\lambda_1 = \lambda_2$ (or $\lambda_1 = - \lambda_2$)}

\vskip 0.5cm

In the case that $\lambda_1 = \lambda_2$ there are additional
generators which are invariant under the conjugation with $t$. Indeed,
taking for example 
\begin{equation}
\alpha = \left( 0,0,0,0,0,0,-1,-1\right) \,\,\, \mbox{and} \,\,\, \beta =
\left( 0,0,0,0,1,1,0,0\right)
\end{equation}
one can show that
\begin{equation} \label{eq:invcomb}
t\left( E_{\alpha} - E_{\beta}\right) t^{-1} = E_{\alpha} - E_{\beta}
, 
\end{equation}
details of this computation can be found in the appendix \ref{ap:e6}.
Analogous combinations exist if we simultaneously permute the last four
entries in $\alpha$ and $\beta$. This adds six generators to
SU(4)$\times$U(1) which yields a 22 dimensional gauge symmetry. This
is consistent with the findings of \cite{Forste:2005rs} where this
gauge group has been identified as SO(7)$\times$U(1). (Since now we
have also superpositions of root operators computing the Dynkin diagram
would involve diagonalising the adjoint action of Cartan
generators. We do not carry out this involved calculation here, the
techniques are described in \cite{Forste:2005rs}. What one does see is
the rank reduction by two since two Cartan generators have been
projected out by the Wilson line.)

For $\lambda_1 = -\lambda_2$ one replaces $E_\alpha -E_\beta$ by
$E_\alpha + E_\beta$ in (\ref{eq:invcomb}). Then the rest of the
discussion is the same as in the case $\lambda_1 = \lambda_2$.

\vskip 1cm

\centerline{\it Special Case: $\lambda_1 \not= 0$, $\lambda_2 =
  0$ (or $\lambda_1 =0$,  $\lambda_2\not= 0$)}

\vskip 0.5cm
 
For $\lambda_2 = 0$ the following SO(10)$\times$U(1) generators are
invariant under conjugation with $t$:
\begin{itemize}
\item five Cartan generators: $H_1 -H_2$, $H_1 -H_3$, $H_1 - H_4$,
  $H_1-H_5$, $H_1 - H_6$,
\item 20 root generators corresponding to root vectors of the form
$\left( 0,0,0,\underline{1,-1,0,0,0}\right)$.
\end{itemize}
The unbroken gauge symmetry can be identified
as SU(5)$\times$U(1). The decoupled U(1) is generated by
\begin{equation} \label{eq:su5u1}
5H_1 - H_2 -H_3 -H_4 -H_5 -H_6 .
\end{equation}
The case $\lambda_1=0$, $\lambda_2 \not= 0$
yields the same result after some obvious sign changes have been
performed. 

The possible breakings due to continuous Wilson lines are summarised
in table \ref{tab:sumcb}

\begin{table}[h]
\centering
\begin{tabular}{| c | c |}
\hline 
choice for parameters in (\ref{eq:genconwile6}) & unbroken gauge group \\
\hline \hline
$\lambda_1 \not=0$, $\lambda_2 \not= 0$, $\lambda_1 \not= \pm
\lambda_2$ &  SU(4)$\times$U(1) \\ \hline
$\lambda_1 = \pm \lambda_2 \not= 0$ & SO(7)$\times$U(1) \\ \hline
$\lambda_1 \not= 0 $, $\lambda_2  = 0$ (or $\lambda_1 = 0$, $\lambda_2
\not= 0$) & SU(5)$\times$U(1)\\ \hline
\end{tabular}
\caption{All symmetry breakings of SO(10)$\times$U(1) by the
  continuous Wilson line.}\label{tab:sumcb}
\end{table}

\subsubsection{The PS to SM Breaking} 

In order to complete the rederivation of the model in
\cite{Forste:2005rs} we should include the fifth direction into the
discussion. Along that direction we have a discrete Wilson line
\begin{equation}
a_5 = \left( 1/2,1/2, -1, 0, 0, 0,-1/2, 1/2\right).
\end{equation}
The projection at the fixed point $x^5 = \pi R^5$, $x^6 = 0$ is
obtained from the shift embedding with shift vector $V+a_5$. The E$_6$
root vectors with an integer valued scalar product with $V+a_5$ are
\begin{itemize}
\item 12 root vectors of the form $\left( 0,0,0, \underline{\pm 1, \pm
  1, 0}, 0,0\right) $ ,
\item four root vectors of the form $\left( 0,0,0,0,0,0, \pm 1, \pm
  1\right)$, 
\item 16 root vectors of the form $\left( \pm \left( \frac{1}{2},
  \frac{1}{2}, \frac{1}{2}\right), \pm \frac{1}{2}, \pm \frac{1}{2},
  \pm \frac{1}{2}, \pm \left( \frac{1}{2}, -\frac{1}{2}\right)\right)$
  with an even number of minus signs.
\end{itemize}
The unbroken gauge group at the point $x^5 = \pi R_5$,
$x^6 =0$ is SU(6)$\times$SU(2). For vanishing continuous Wilson line,
the four dimensional gauge symmetry is obtained by imposing in
addition integer valued scalar products with $V$. This removes the
spinorial roots and leads to
SU(4)$\times$SU(2)$\times$SU(2)$\times$U(1), i.e.\ the Pati-Salam
group and an extra U(1). In summary, E$_6$ is broken to the Pati-Salam
group and an extra U(1) at the compactification scale. 

\begin{figure}[h!]
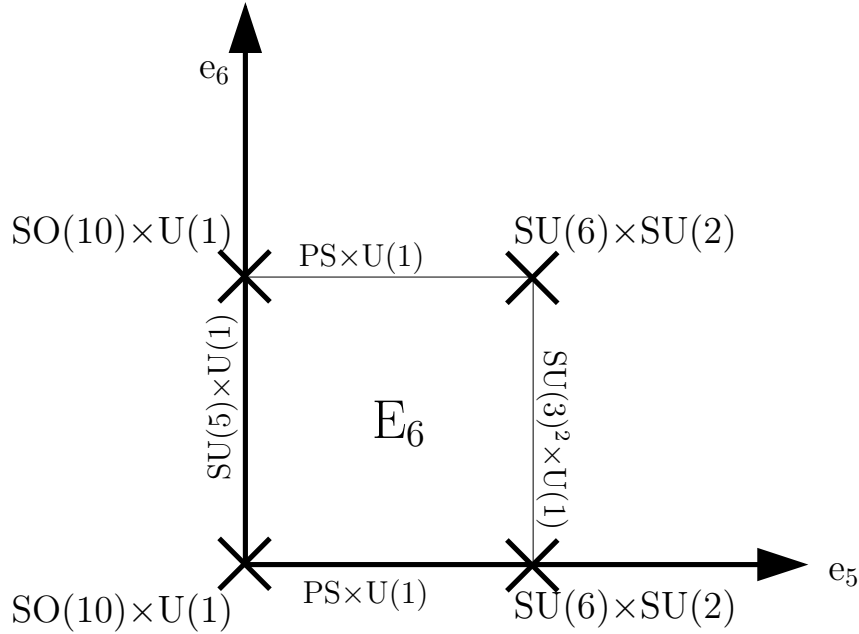

\begin{center}\input gaugegroup_geography.pstex_t \end{center}
\caption{Local projections of the adjoint of E$_6$ to adjoints of
  groups written at the fixed points. Groups written at lines
  connecting two fixed points are obtained after applying both
  projections. The unbroken gauge symmetry in four dimensions is
  SU(3)$\times$SU(2)$\times$U(1)$^2$. } 
\label{fig:gaugegroupgeography}
\end{figure}

A continuous Wilson line can be used to break that symmetry further. 
We focus on the $x^6$ direction with the continuous Wilson line 
\begin{equation}
T = \lambda \left( E_\delta + E_{-\delta}\right) ,
\end{equation}
and $\delta$ given by (\ref{eq:delta}). The background fieldstrength
is zero since $\delta \cdot a_5 =0$ (the discrete and continuous
Wilson lines commute). According to our general discussion the
unbroken gauge group at $\left( 0, \pi R_6 \right)$ is
SO(10)$\times$U(1) and at $\left( \pi R_5 , \pi R_6\right)$
SU(6)$\times$SU(2). The intersection of the unbroken groups at
$\left( 0,\pi R_6 \right)$ and $\left( \pi R_5 , \pi R_6\right)$ is
again SU(4)$\times$SU(2)$\times$SU(2)$\times$U(1). The intersection 
of groups at the points $\left( 0,0\right)$ and $\left( 0, \pi R_6
\right)$ was determined in section \ref{sec:wlbreaking} to be
SU(5)$\times$U(1). It remains to compute the intersection at $\left(
\pi R_5 , 0\right)$ and $\left( \pi R_5 , \pi R_6\right)$ and the
intersection of all fixed point groups giving the unbroken four
dimensional symmetry. Let us first discuss the points  $\left(
\pi R_5 , 0 \right)$ and $\left( \pi R_5 , \pi R_6\right)$. We start
e.g.\ with the SU(6)$\times$SU(2) at  $\left( \pi R_5 , 0\right)$ and
keep only generators which are invariant under conjugation with
$t$. These are:
\begin{itemize}
\item five Cartan generators: $H_1 -H_2$, $H_1 - H_3$, $H_1 - H_4$,
  $H_1 - H_5$ and $H_1 - H_6$ ,
\item six root generators with root vectors of the form $ \left(
  0,0,0,\underline{1,-1,0},0,0\right)$ ,
\item two root generators of the form $\left( 0,0,0,0,0,0,
  \underline{1,-1}\right) $,
\item four root generators with root vectors of the form
$\left( \pm\left(\frac{1}{2}, \frac{1}{2}, \frac{1}{2}\right),
  \pm\left( \frac{1}{2},
  \frac{1}{2}, \frac{1}{2}\right), \pm\left( \frac{1}{2},
  -\frac{1}{2}\right) \right) $ with an even number of minus signs.
\end{itemize}
Working out the details of the algebra one finds for
the intersection of the gauge groups unbroken at  $\left(
\pi R_5 , 0\right)$ and $\left( \pi R_5 , \pi R_6\right)$ the group
SU(3)$\times$SU(3)$\times$U(1), where the U(1) factor is generated by
\begin{equation} \label{eq:firstu1}
 -3H_1 - H_2 -H_3 -H_4 +3H_5 +3H_6 .
\end{equation}
The four dimensional unbroken gauge group can now be obtained by e.g.\
keeping only those SU(3)$\times$SU(3)$\times$U(1) root generators
whose root vectors have an integer valued scalar product with the
shift vector $V$ (\ref{eq:rabyshift}). This removes the spinorial
roots, the second SU(3) is broken to SU(2)$\times$U(1) where the U(1)
generator is given by
\begin{equation} \label{eq:secondu1}
3H_1 -H_2 -H_3 -H_4 .
\end{equation}
The unbroken gauge group in four dimensions is
SU(3)$\times$SU(2)$\times$U(1)$\times$U(1)  
where the two U(1)s are generated by (\ref{eq:firstu1}) and
(\ref{eq:secondu1}). Note, that the charge 
vectors are orthogonal
\begin{equation}
\begin{array}{lll}
q_1 &=& \left( -1,-1,-1,-1,-1,-1,3,3\right)\\
q_2 & =& \left( 1,1,1, -1,-1,-1,0,0\right)
\end{array} \to \, q_1 \cdot q_2 = 0.
\end{equation}
Now, we have completely
rederived the picture drawn in figure 2 of \cite{Forste:2005rs} (see
figure \ref{fig:gaugegroupgeography}).  

\subsection{Splitting Bulk Matter} \label{sec:splitting}

Bulk matter appears in complete representations of the bulk group and
splits under the orbifold projections into representations of the
unbroken gauge group at a particular fixed point. Our previous
calculations were simple because in most cases we needed only to know
whether a commutator is zero or something else. For that one does not
have to determine all the structure constants but just needs to check
whether two roots ad up to another root which is much simpler. In
order to carry over that degree of simplicity also to the computation
of bulk matter projections it is useful to view gauge and matter
fields as part of the adjoint representation of E$_8$. Then one needs
to know only whether the sum of two E$_8$ roots yields another E$_8$
root. 

We view the E$_6$ as a
factor of an E$_6\times$U(1)$\times$U(1) subgroup of E$_8$. 
We introduce two additional Cartans into (\ref{eq:eightcart})
\begin{equation}\label{eq:e8nonstan}
\left( H_1 + K, H_1 + L, H_1 - K -L, H_2, H_3,H_4, H_5, H_6\right) .
\end{equation}
The generators $K$ and $L$ are two additional U(1) generators which
are not necessarily related to symmetries. However, as will be seen
shortly, charges under these additional U(1)s provide a useful tool
for identifying E$_6$ representations within the set of E$_8$
generators.  
The E$_8$ root vectors are
\begin{itemize}  
\item 112 vectors of the form $\left( \underline{\pm 1 , \pm 1,
  0,0,0,0,0,0}\right)$ ,
\item 128 vectors of the form $\left( \pm \frac{1}{2},\pm
  \frac{1}{2},\pm  \frac{1}{2},\pm \frac{1}{2},\pm \frac{1}{2}, \pm
  \frac{1}{2},\pm \frac{1}{2},\pm \frac{1}{2}\right)$ with an even
  number of minus signs.
\end{itemize}
Root operators which differ from our previously used E$_6$ root
operators fall into E$_6\times$U(1)$^2$ representations. In order to
find an E$_6$ representation we collect elements with identical $K$
and $L$ charges, e.\ g.\ for $K=2/3$, $L=-1/3$ one finds 27 root
generators with the following root vectors:
\begin{itemize}
\item one vector with $H_1 = -2/3$: $\left( 0,-1,-1,0,0,0,0,0\right) ,$
\item ten vectors with $H_1 = 1/3$: $\left( 1,0,0,\underline{\pm
  1,0,0,0,0}\right) ,$
\item 16 vectors with $H_1 = -1/6$: $\left( \frac{1}{2},
  -\frac{1}{2},-\frac{1}{2}, 
  \pm \frac{1}{2}, \pm \frac{1}{2}, \pm \frac{1}{2}, \pm \frac{1}{2},
  \pm \frac{1}{2}\right)$ (even number of minus signs).
\end{itemize}
That this is indeed the ${\bf 27}$ of E$_6$ can be established by
computing the Dynkin labels. Since a detailed discussion on how to do
this can be found in the literature (e.g.\cite{Giedt:2001zw,Akin}) we
refrain from the presentation of the relevant calculation.
The {\bf $\overline{\mbox{\bf 27}}$} can be obtained by
multiplying the first three entries of the 
root vectors appearing in the {\bf 27} with $-1$ (reversing the
$\left( H_1 , K, L\right)$ charges). There are two more pairs of {\bf
  27}, {\bf $\overline{\mbox{\bf 27}}$} 
in the branching of E$_8$
which can be easily obtained by permuting the first three entries of
our set. (The remaining six generators with root vectors of the form 
$\left(\underline{1,-1,0},0,0,0,0,0\right)$ are E$_6$ singlets.)

In orbifold field theory constructions one usually chooses orbifold
parities for matter by hand. These can be different for different
fixed points. If we choose opposite parities for
points differing only 
in the $x^6$ direction the complete E$_6$ multiplet will be projected
out. Therefore we take identical parities for those points and it is
enough to characterise the assigned parities by an 
ordered pair $\left( \pm , \pm \right)$ corresponding to the values at
fixed points with $x^5 =0$, $x^5=\pi R_5$, respectively.

After we embedded everything into the adjoint of E$_8$ the technical
details of computing the local projections are the same as for the
gauge group. We omit their presentation here, and just report the
results. Further, we restrict our discussion to a {\bf 27}. The
$\overline{\mbox{\bf 
    27}}$ gives rise to charge conjugated states.

At the fixed point $(0,0)$ the gauge symmetry is SO(10)$\times$U(1)
and the possible splittings of a {\bf 27} are
\begin{equation}
\begin{array}{l l }
\mbox{\bf 27} \to \mbox{\bf 10}_{1/3} + \mbox{\bf 1}_{-2/3}
&  \mbox{for positive parity} ,\\
\mbox{\bf 27} \to  \overline{\mbox{\bf 16}}_{-1/6} 
&  \mbox{for negative parity.}
\end{array}
\end{equation}
At the fixed point $\left( \pi R_5 , 0\right)$ the bulk group is
broken to SU(6)$\times$SU(2) and the {\bf 27} splits into 
\begin{equation}
\begin{array}{l l }
\mbox{\bf 27} \to \mbox{\bf ($\overline{\mbox{\bf 6}}$,2)} &
\mbox{for positive parity} ,\\ 
\mbox{\bf 27} \to  \mbox{\bf (15,1)} &  \mbox{for negative parity.} 
\end{array}
\end{equation}
The intersection of gauge groups at horizontally separated fixed
points is SU(4)$\times$\-SU(2)$\times$\-SU(2)$\times$\-U(1). Depending on the
assigned parities bulk matter splits according to
\begin{equation}
\begin{array}{l l }
\mbox{\bf 27} \to \mbox{\bf (1,2,2)}_{1/3}
&  \mbox{for $\left( + ,+\right)$ parity} ,\\
\mbox{\bf 27} \to  \mbox{\bf (1,1,1)}_{-2/3} + \mbox{\bf (6,1,1)}_{1/3}
&  \mbox{for $\left( + , -\right)$ parity,}\\
\mbox{\bf 27} \to  \mbox{\bf ($\overline{\mbox{\bf 4}}$,2,1)}_{-1/6}
&   \mbox{for $\left( - , +\right)$ parity,}\\
\mbox{\bf 27} \to  \mbox{\bf (4,1,2)}_{-1/6}
&  \mbox{for $\left( - , -\right)$ parity.}
\end{array}
\end{equation}

Following our general discussion the gauge groups and matter
spectrum at vertically separated fixed point are the same but the
embedding into the bulk gauge group and matter differs. For $x^5 = 0$
the intersection is SU(5)$\times$U(1) and the local projections of the
bulk {\bf 27} intersect in
\begin{equation}\label{eq:sosu}
\begin{array}{l l }
\mbox{\bf 27} \to \overline{\mbox{\bf 5}}_{8/3}   &  \mbox{for
  positive parity} ,\\ 
\mbox{\bf 27} \to \overline{ \mbox{\bf 10}}_{-4/3}  &  \mbox{for negative
  parity.} 
\end{array}
\end{equation}
Accordingly for $x^5 = \pi R_5$ the intersection of the two
SU(6)$\times$SU(2) is SU(3)$\times$SU(3)$\times$U(1) and the
intersections for the matter projections are
\begin{equation}
\begin{array}{ll}
\mbox{\bf 27} \to \mbox{\bf (1,$\overline{\mbox{\bf 3}}$)}_{-2} +
  \mbox{\bf (3,1)}_{2}   &  \mbox{for positive parity} ,\\
\mbox{\bf 27} \to  \mbox{\bf($\overline{\mbox{\bf
      3}}$,3)}_0  &  \mbox{for negative parity.}
\end{array}
\end{equation}
Finally, the massless fields in four dimensions are given as the
intersections from all fixed points. One obtains the following
SU(3)$\times$\-SU(2)$\times$\-U(1)$\times$\-U(1) multiplets
\begin{equation}
\begin{array}{ll}
\mbox{\bf 27} \to \mbox{\bf (1,2)}_{-2,1} & \mbox{for $\left(
  +,+\right)$ parity},\\
\mbox{\bf 27} \to \mbox{\bf ($\overline{\mbox{\bf
  3}}$,1)}_{0,2} & \mbox{for $\left( +,-\right)$ parity},\\
\mbox{\bf 27} \to  {\bf (1,1)}_{-2, -2} + {\bf
  (3,1)}_{2, 0} & \mbox{for $\left( -,+\right)$ parity},
  \\
\mbox{\bf 27} \to \mbox{\bf ($\overline{\mbox{\bf
  3}}$,2)}_{0,-1} & \mbox{for $\left( -,-\right)$ parity}.
\end{array}
\end{equation}
Note, that counting the number of states surviving projections
with any parity gives 15 and not 27. The continuous Wilson line
removes states independent of the assigned parity. Effectively these
states become massive via the coupling to an internal gauge field
component which acquires a VEV. We were able to derive the symmetry
breaking patterns without any explicit computation of Yukawa
couplings. The reason for this is that couplings between bulk matter
and the continuous Wilson line are completely fixed by higher
dimensional gauge invariance.

\begin{figure}[h!]
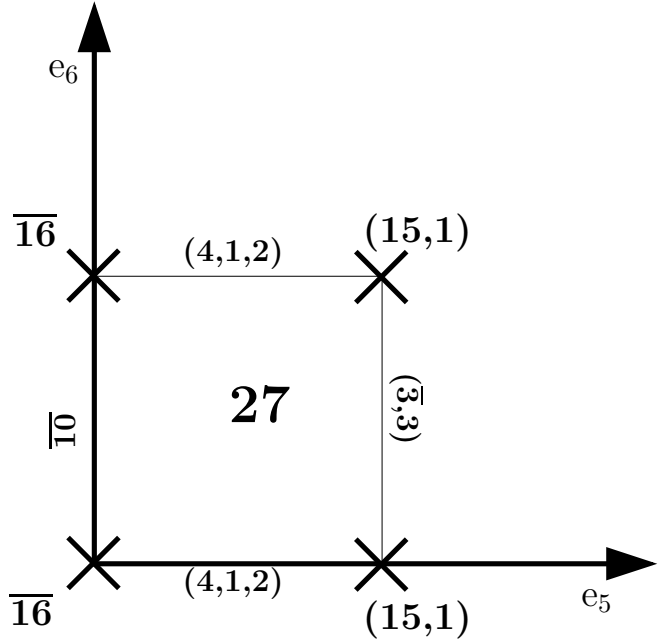

\begin{center}\input matter_geography.pstex_t \end{center}
\caption{Local projections of the {\bf 27}. At all fixed points
  negative parity has been assigned. States written at lines
  connecting two fixed points survive both projections. The massless
  state in four dimensions is {\bf ($\overline{\mbox{\bf
  3}}$,2)}.}
\label{fig:mattergeography}
\end{figure}

We are now in a position to draw figure \ref{fig:gaugegroupgeography} for
bulk matter. We do this in figure \ref{fig:mattergeography} with a {\bf
  27} in the bulk and 
assigned parity $\left( - ,-\right)$.
\section{Electroweak Symmetry Breaking} \label{sec:ewstrat}

In the rest of the paper we will explore the
possibility to assign the electroweak symmetry breaking to continuous
Wilson lines. 

As an illustrating example we
focus on models where an SO(10) gauge symmetry appears at an
intermediate level in the breaking of the heterotic E$_8 \times$E$_8$
gauge symmetry to the Standard Model symmetry in four dimensions. As
argued in \cite{Forste:2004ie,Nilles:2004ej} such models can be
phenomenologically favourable since some of the appealing features in
SO(10) GUT theories are inherited even if SO(10) is broken in four
dimensions. 

In the example of the previous section such an intermediate SO(10)
appeared at the origin of the extra dimensions torus lattice. This
example is, however, not suitable in the context of electroweak symmetry
breaking by continuous Wilson lines. The 78 dimensional adjoint
representation splits into the 45 dimensional adjoint of SO(10), a
{\bf 16}, a $\overline{\mbox{\bf 16}}$ and a singlet of SO(10). Hence
the continuous Wilson lines transform in 16 dimensional
representations whereas the Standard Model Higgs should be embedded
into a ten dimensional representation of SO(10). From that example we
learn what we have to look for. The SO(10) should come from a broken
group $G$ whose adjoint branches into the {\bf 45} and {\bf 10} and
possibly other SO(10) representations. In such models a continuous
Wilson line transforming in the {\bf 10} exists. Under the subsequent
breaking to the Standard Model gauge group the {\bf 10} can split such
that only an SU(2) doublet remains massless. This would give rise to
a continuous Wilson line corresponding to the Standard Model Higgs doublet.
(In the models considered later there will be more than one {\bf 10}
giving rise to more doublets in four dimensions. Among those doublets
there will be always the pair of the MSSM.)

In the following two sections we will present two examples where these
ideas are realised to some extent. The intermediate SO(10) will arise
from a ${\mathbb Z}_2$ breaking of SO(14) and SO(12). The branching
rules for such breakings fall into the category discussed above.

\section{A ${\mathbb Z}_6$-II Example}\label{eq:z6ex}

In this section we consider a heterotic E$_8 \times$E$_8$ ${\mathbb
  Z}_6$-II orbifold. The group ${\mathbb Z}_6$ is isomorphic to
  ${\mathbb Z}_3 \times {\mathbb Z}_2$. The embedding of ${\mathbb
  Z}_3$ will be taken such that the visible E$_8$ is broken to
  SO(14)$\times$U(1). The subsequent ${\mathbb Z}_2$ breaking to SO(10) will
  give rise to continuous Wilson lines in ten dimensional
  representations of SO(10). A combination of discrete and continuous
  Wilson lines will be applied to finally break the symmetry to the
  colour SU(3) the electromagnetic U(1) and extra U(1)s. Here, we
  assume that the three families of chiral matter originate from 16
  dimensional representations of the intermediate SO(10). In the
  following, we will
  separate the discussion of matter from our studies and focus on the
  gauge sector, only. 

\begin{figure}[h!]
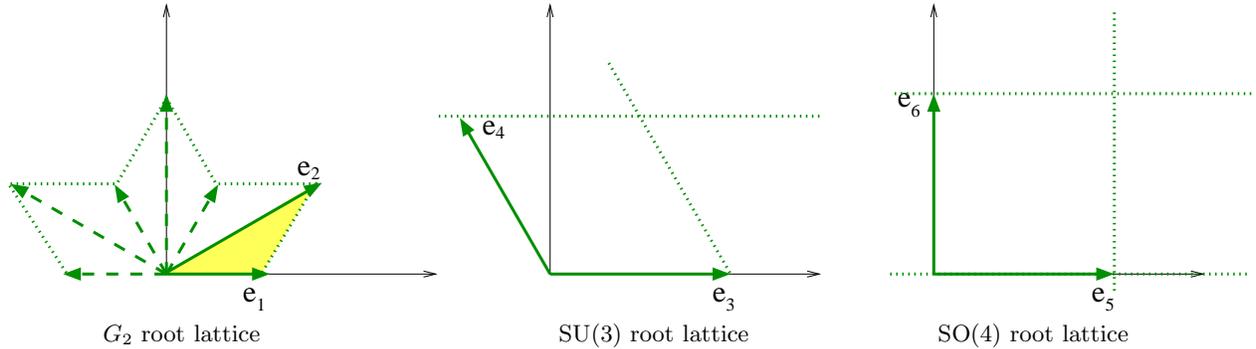

\centering
\begin{center}\input z6-geometry.pstex_t \end{center}
\caption{Geometry of compact space.}
\label{fig:z6geom}
\end{figure}

Before describing the details of gauge symmetry breaking we recall the
structure of the six dimensional space the heterotic string is
compactified on (see e.g.\ \cite{Kobayashi:2004ud}). This is taken as
a ${\mathbb Z}_6$-II orbifold of a 
six dimensional torus. The six dimensional torus is a product of three
two dimensional tori whose compactification lattices are the root
lattices of G$_2$, SU(3) and SO(4), respectively. 
In each of the torus lattices the orbifold acts as a rotation.
The action of
${\mathbb Z}_6$ is characterised by a three component vector whose
entries are rotation angles in $2\pi$ units. For ${\mathbb Z}_6$-II
this vector is
\begin{equation} \label{eq:z6IIspace}
v = \left( \frac{1}{6}, \frac{1}{3}, -\frac{1}{2}\right) .
\end{equation}
One characteristic feature of ${\mathbb Z}_6$-II is that ${\mathbb
  Z}_3$ as well as ${\mathbb Z}_2$ posses fixed tori. The structure of
  the compact space is visualised in figure \ref{fig:z6geom}.
 
\subsection{Rotational Embedding}

In this section we discuss the model within the rotational
embedding approach of \cite{Ibanez:1987xa}. We will focus only on the
first E$_8$. In the next section, we will consider an equivalent model
where the orbifold is embedded as a shift. For the equivalent model we
will also give the components of shift vectors and discrete Wilson
lines within the second E$_8$ such that modular invariance is
ensured. 

The orbifold group is ${\mathbb Z}_6 = {\mathbb Z}_3 \times {\mathbb
  Z}_2$. We embed the ${\mathbb Z}_3$ factor as a shift and the
  ${\mathbb Z}_2$ factor as a rotation. The ${\mathbb Z}_3$ shift is
  taken to be\footnote{Compared to (\ref{eq:e8nonstan}) we use the
  canonical E$_8$ Cartan vector $\left( H_1, H_2 , H_3 , H_4, H_5,
  H_6, H_7 , H_8\right)$ in the rest of the paper.} 
\begin{equation}
V_3 = \left( \frac{4}{3}, 0^7\right) ,
\end{equation}
where $0^n$ is shorthand for $n$ vanishing consecutive components. 
E$_8$ roots having integer valued scalar product with $V_3$ are
\begin{equation}
\left( 0, \underline{ \pm 1, \pm 1, 0^5}\right) .
\end{equation}
Hence, ${\mathbb Z}_3$ breaks E$_8$ to SO(14)$\times$U(1). 
The ${\mathbb Z}_2$ is embedded as a rotation acting on the 16 extra
bosonic chiral fields $X^I$ of heterotic string theory which are
compactified on an E$_8\times$E$_8$ lattice.  This induces a rotation
acting on Cartan vectors ($\sim \partial X^I$). Root vectors are
charge vectors for the adjoint representation and are rotated
accordingly. Consistency requires that root vectors are mapped to root
vectors or in other words that the rotation is an automorphism of the
E$_8 \times$E$_8$ root lattice. For details see
\cite{Ibanez:1987xa,Forste:2005rs}. We take the following SO(8) matrix
for the rotational embedding of ${\mathbb Z}_2$ within the first E$_8$
\begin{equation}\label{eq:orpirot}
s = \mbox{diag}\left( 1,1,1,1, -1,-1,-1,-1\right) .
\end{equation}
Actually, it is not enough to define an automorphism on the root
lattice. Indeed, one should specify an automorphism on the E$_8$ Lie
algebra. We will come back to that later and  for the time being keep
the discussion at a 
schematic level. For now, we will also just
compute the rank and dimension of unbroken gauge groups and guess the
algebra. All these results have been computed in a rigorous
by realising the gauge embedding as an algebra automorphism. 
For presentational purposes the sketchy derivations given at the
moment are more suitable. 

The Cartan operators which are invariant under a rotation by $s$ are
\begin{equation}\label{eq:fourcartans}
H_1,\ldots , H_4. 
\end{equation}
There are 12 invariant roots of SO(14) of the form
$$\left( 0, \underline{ \pm 1, \pm 1, 0}, 0^4\right) . $$
For the remaining 72 roots a superposition of a root operator with its
image is invariant. That leads to 36 more generators for the unbroken
gauge group. So, altogether the unbroken gauge group has dimension
52. The orbifold alone does not reduce the rank of the gauge
group. The four Cartan generators which have been projected
out can be replaced by the invariant combinations
\begin{equation}\label{eq:newcg}
E_{\gamma_i} + E_{-\gamma_i},\,\,\, i= 1, \ldots , 4
\end{equation}
with
\begin{equation}\label{eq:rootsfornwc}
\gamma_1 = \left(0^4, 1, 1, 0, 0\right), \,\,\,  \gamma_2 =\left(0^4,
    1, -1, 0, 0\right) , \,\,\, 
    \gamma_3 = \left(0^6, 1, 1\right) ,\,\,\, \gamma_4 = \left( 0^6,
    1, -1 \right) .
\end{equation}
Indeed, these operators mutually commute ($\gamma_i \pm \gamma_j$ is
not an E$_8$ root) and are neutral under the adjoint action with the
four Cartans in (\ref{eq:fourcartans}). So, the unbroken gauge group has
rank 8 and dimension 52 which fits with\footnote{The rotational
  embedding (\ref{eq:orpirot}) breaks E$_8$ to E$_7 \times$SU(2)
  \cite{Akin}.} 
\begin{equation}\label{eq:intmetesoten}
\mbox{SO(10)}\times\mbox{SU(2)}\times\mbox{SU(2)}\times\mbox{U(1)} ,
\end{equation}
where the extra U(1) factor is the same as in the previous ${\mathbb
  Z}_3$ breaking of E$_8$ to SO(14)$\times$U(1). The ${\mathbb Z}_2$
  breaking of SO(14) to SO(10) falls into the class of models
  considered in the previous section. Hence, continuous Wilson lines
  of the form
\begin{equation}\label{eq:cwlbeforedwl}
W = \sum_{i=5}^8 \lambda_i H_i
\end{equation}
are of potential interest for electroweak symmetry breaking. These
Wilson lines can correspond to shifts by $e_5$ or $e_6$. (For details
on continuous Wilson lines in rotational embeddings see
\cite{Ibanez:1987xa,Forste:2005rs}.) 
Before we study the continuous
Wilson line breaking we will break SO(10) to the Standard Model gauge
group with discrete Wilson lines.  
				   
First, we introduce ${\mathbb Z}_3$ discrete Wilson lines.
Shifts along $e_3$ and $e_4$ will be associated with a shift in
E$_8$. The shift vector is chosen as
\begin{equation}\label{eq:discz3}
a_3= a_4 = \left( 0, \frac{2}{3}, -\frac{1}{3}, -\frac{1}{3}, 0^4\right) .
\end{equation}
In the presence of that Wilson line, the ${\mathbb Z}_3$ sector of the
orbifold breaks more gauge symmetry, E$_8$ root vectors having integer
valued scalar products with $V_3$ and $a_3$ are of the form
\begin{equation}\label{eq:3brech}
\begin{array}{r l l}
\left( 0, \underline{1,-1,0}, 0^4\right) & \rightarrow & \mbox{SU(3)},\\
\left( 0^4 ,\underline{ \pm 1, \pm 1, 0,0}\right) & \rightarrow &
\mbox{SO(8)} ,
\end{array} 
\end{equation}
and the previously unbroken SO(14)$\times$U(1) is broken further to
\begin{equation}
\mbox{SU(3)}\times\mbox{SO(8)}\times\mbox{U(1)}^2 
\end{equation}
by the Wilson line (\ref{eq:discz3}). Breaking this further down by
the ${\mathbb Z}_2$ orbifold leads to
SU(3)$\times$\-SU(2)$^4\times$\-U(1)$^2$ unbroken symmetry in four
dimensions. 

We can obtain a smaller four dimensional gauge group by
introducing an additional ${\mathbb Z}_2$ discrete Wilson line. To
this end, we associate shifts by $e_5$ with a rotation in the E$_8$
root lattice. That rotation describes a discrete Wilson line if it
commutes with the orbifold rotation $s$ in (\ref{eq:orpirot}). We take
\begin{equation}\label{eq:dwlrot}
a_5 = \mbox{diag}\left( 1,1,1,1, -\sigma , -\sigma\right), \,\,\, \mbox{with}
\,\,\, \sigma = \left( \begin{array}{cc} 0 & 1 \\ 1 & 0 \end{array}
\right) . 
\end{equation}
To our knowledge, this option of describing discrete Wilson lines in
rotationally embedded orbifolds has not been considered in the
literature, so far. Later we will argue that in a basis where the
orbifold becomes shift embedded this Wilson line corresponds to
a discrete Wilson line as introduced in \cite{Ibanez:1986tp}. 

Generators for the unbroken gauge group in four dimensions are
obtained by imposing invariance under $s$ and $a_5$ rotations on the
generators of SU(3)$\times$\-SU(2)$^4\times$\-U(1)$^2$. Again, the four
Cartan operators (\ref{eq:fourcartans}) are invariant.
The six SU(3) roots in (\ref{eq:3brech}) are also invariant under
orbifold and Wilson line. 
The structure of invariant superpositions of root operators is a bit
more complex than previously. Some orbifold invariant superpositions
are also invariant under the Wilson line. This happens if either the
involved roots are invariant under the rotation by $a_5$ or the $s$
and $a_5$ image of a root are identical. This applies to the four new
Cartan generators in (\ref{eq:newcg}) and we see that the discrete
Wilson line does not reduce the rank of the gauge group. The remaining
$24-8=16$ roots of SO(8) lead to four invariant superpositions of four
root generators. Altogether, the rank of the unbroken gauge group is
eight and the dimension is 18. This fits with 
$$ \mbox{SU(3)}\times\mbox{SU(2)}\times \mbox{U(1)} \times\mbox{SU(2)}
\times \mbox{U(1)}^3 .$$
This result has been confirmed by analysing the algebra of the
unbroken gauge group as we will discuss below. Such an analysis shows
also that an SU(3)$\times$SU(2)$\times$U(1) factor is completely
embedded into the intermediate SO(10) in (\ref{eq:intmetesoten}). We
interpret that subgroup as the Standard Model group.  

Let us now study the symmetry breaking by a continuous Wilson
line. The continuous Wilson line we are going to consider appears as a
non trivial embedding of a lattice shift by $e_6$. 
The embedding of the discrete Wilson line as a rotation provides
a novel feature, {\it viz.} the stabilisation of some parameters in
the continuous Wilson line (\ref{eq:cwlbeforedwl}). Shifts by $e_5$
and $e_6$ should commute and hence the continuous Wilson line can only
point into directions which are left invariant under the rotation
$a_5$ in (\ref{eq:dwlrot}). That reduces the number of four parameters
in (\ref{eq:cwlbeforedwl}) to two
\begin{equation}
W = \lambda \left( H_5 - H_6 \right) + \lambda^\prime \left( H_7 - H_8
\right) .
\end{equation}
The effective four dimensional picture for such a moduli stabilisation
is that a non vanishing fieldstrength $F_{56}$ results in mass terms for
some of the continuous Wilson line moduli. Since the stabilisation
mechanism is due to a discrete Wilson line the associated scale is the
compactification scale. 

\begin{figure}[h!]
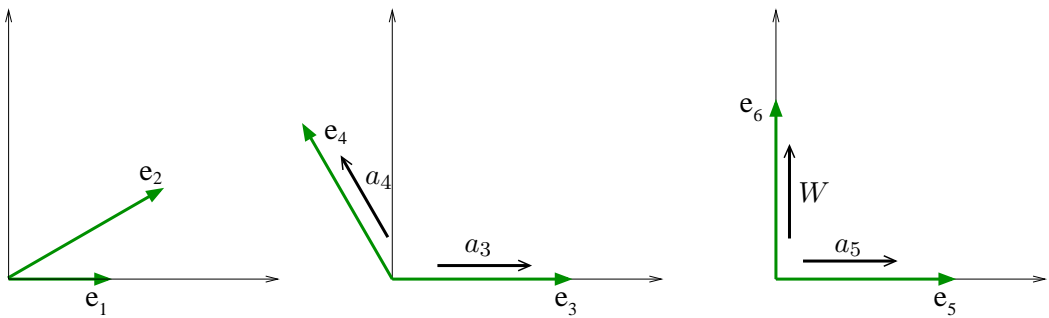

\centering
\begin{center}\input z6-geometry-with-wl.pstex_t \end{center}
\caption{The setup of the model. The Wilson lines $a_3$, $a_4$, and $a_5$
are {\it discrete}, where the first two, $a_3$ and $a_4$, are realised as
shifts, and the latter, $a_5$, as a rotation. $W$ is {\it continuous}. The
geometry does not allow for Wilson lines in the first torus.}
\label{fig:z6_model_setup_wilsonlines}
\end{figure}

Let us focus on the example $\lambda \not= 0$ and $\lambda^\prime =
0$. In this case, generators which are charged under $H_5 - H_6$ are
projected out.  
This removes the Cartan operator $E_{\gamma_2} + E_{-\gamma_2}$ from
the list in (\ref{eq:newcg}) and the rank is reduced by one. All
quadruplets forming superposition invariant under $s$ and $a_5$ are
projected out as well. Hence, the unbroken gauge group is
\begin{equation}
\mbox{SU(3)} \times \mbox{U(1)}^5 .
\end{equation}
That this breaking corresponds indeed to electroweak symmetry breaking
as far as the Standard Model subgroup of the intermediate SO(10) is
concerned can be confirmed by a detailed analysis of the Lie
algebras. The distribution of all appearing Wilson lines within the
compact space is summarised in figure
\ref{fig:z6_model_setup_wilsonlines}. 

After we presented the symmetry breaking pattern at a schematic level
we put the derivations on firm ground now. To this end, we need to
provide the orbifold embedding as an algebra automorphism and work out
the algebras of the unbroken gauge symmetries. We employ the same
techniques as we did in \cite{Forste:2005rs} where also more details
can be found. 

So far the ${\mathbb Z}_2$ orbifold and Wilson line embeddings have
been given as automorphisms of the E$_8$ root lattice in
(\ref{eq:orpirot}) and (\ref{eq:dwlrot}), respectively. As such they
can be written as a product of Weyl reflections. The orbifold
embedding (\ref{eq:orpirot}) can be obtained as the product of four Weyl
reflections at $\gamma_1$ to $\gamma_4$ in
(\ref{eq:rootsfornwc}). Methods on how to obtain an algebra
automorphism from Weyl reflections are discussed e.g.\ in
\cite{Schellekens:1987ij,Bouwknegt:1988hn}. A canonical guess for the
algebra automorphism belonging to the orbifold action
(\ref{eq:orpirot}) is the conjugation with
\begin{equation}\label{eq:shiftlesslift}
e^{\frac{i\pi}{2}\sum_{j=1}^4\left( E_{\gamma_j} +
  E_{-\gamma_j}\right)} ,
\end{equation}
where the appearing operators are the same as given in
(\ref{eq:newcg}). Next, one needs to ensure that the algebra
automorphism represents a ${\mathbb Z}_2$. All structure constants of
E$_8$ have to be determined. A systematic way of doing that can be
found in \cite{carter}, see also Appendix \ref{ap:e6}. We used a
computer programme to perform that calculation and found that the
algebra automorphism is indeed ${\mathbb Z}_2$. It corresponds to the
lift of conjugacy class 44 in \cite{Akin}.

The rotation (\ref{eq:dwlrot}) belonging to the discrete Wilson line
can be written as the product of Weyl reflections at
$\gamma_1$ and $\gamma_3$. Here, the situation is slightly more complicated
since
the canonical guess provides a ${\mathbb Z}_4$ automorphism of
E$_8$. As in \cite{Forste:2005rs}, we supplement the canonical guess
with an additional shift. Embedding the Wilson line (\ref{eq:dwlrot})
as a conjugation with
\begin{equation}\label{eq:wllift}
e^{\frac{i\pi}{2}\sum_{j=1}^4 H_j} e^{\frac{i \pi}{2}\left(
   E_{\gamma_1} + E_{-\gamma_1}-E_{\gamma_3} - E_{-\gamma_3}\right)}
\end{equation}
provides a consistent ${\mathbb Z}_2$ representation. The additional
minus sign in front of $E_{\pm \gamma_3}$ is not important for the
consistency of the algebra automorphism but for modular invariance. We
will comment on that shortly.

After the complete embedding into the algebra has been fixed the
unbroken gauge groups can be identified also on an algebraic level. 
We have checked that our previous claims on the gauge symmetry
breaking pattern are correct. The detailed calculations go along
the same lines as the one presented in \cite{Forste:2005rs}. Instead
of reporting them here we will present an equivalent model where
orbifold and discrete Wilson lines are embedded as shifts. That model
will be investigated in detail in the next section by using the
technique developed in section \ref{sec:wl}.

Since all generators appearing in the exponents in
(\ref{eq:shiftlesslift}) and (\ref{eq:wllift}) are elements of the new
Cartan algebra defined in (\ref{eq:newcg}) the conjugation with these
elements can be associated with a shift embedding w.r.t.\ the new
choice of a Cartan subalgebra\footnote{The fact that discrete Wilson
  line and orbifold appear as shifts w.r.t.\ the same Cartan
  subalgebra is a direct consequence of the defining property that
  orbifold embedding and lattice shift embedding commute for discrete
  Wilson lines.}. In order to make contact to standard
notation one has to perform redefinitions within the set of new Cartan
generators such that the E$_8$ roots take their standard form given in
section \ref{sec:splitting}. First, one has to diagonalise the adjoint
action of the operators (\ref{eq:newcg}), or in other words find a
Cartan Weyl basis in which the operators (\ref{eq:newcg}) together
with $H_1,\ldots, H_4$ form the Cartan subalgebra. This has been done
with the help of a computer programme. If the eigenvalues corresponded
to the standard roots of E$_8$ the shift vector could be read of from
(\ref{eq:shiftlesslift}) as $\left( 0^4, \frac{1}{4}, \frac{1}{4},
\frac{1}{4}, \frac{1}{4} \right)$. The actual calculation shows
however, that in order to obtain the E$_8$ roots in their standard
form one has to take e.g.\ the following Cartan subalgebra (denoted by
$\overline{H}_1, \ldots , \overline{H}_8$)
\begin{equation}
\begin{array}{r c l}
H_i & = & \overline{H}_i \,\,\, \mbox{for $i=1, \ldots, 4$}, \\
E_{\gamma_1} + E_{-\gamma_1}& = & \overline{H}_8 - \overline{H}_6 , \\
E_{\gamma_2} + E_{-\gamma_2}& = & \overline{H}_8 + \overline{H}_6 , \\
E_{\gamma_3} + E_{-\gamma_3}& = & \overline{H}_7 - \overline{H}_5 , \\
E_{\gamma_4} + E_{-\gamma_4}& = & \overline{H}_7 + \overline{H}_5.
\end{array}
\end{equation}
Plugging these redefinitions into (\ref{eq:shiftlesslift}) and
(\ref{eq:wllift}) one obtains expressions of the form
\begin{equation}
e^{2 \pi i V \cdot \overline{H}} .
\end{equation}
For the orbifold the shift vector $V$ is
\begin{equation}
V_2 = \left( 0^6, \frac{1}{2}, \frac{1}{2}\right) .
\end{equation}
The Wilson line $a_5$ becomes also shift embedded, and slightly
abusing the notation we call the shift vector $a_5$, too. Taking into
account the additional first term in (\ref{eq:wllift}) we find
\begin{equation}
a_5 = \left( \frac{1}{4}, \frac{1}{4}, \frac{1}{4}, \frac{1}{4},
\frac{1}{4}, -\frac{1}{4}, -\frac{1}{4}, \frac{1}{4} \right) .
\end{equation}
For modular invariance it is necessary that the scalar product of
$V_2$ with $a_5$ is an integer multiple of 1/2. With our sign choice in
(\ref{eq:wllift}) that goal has been achieved. (Another possibility is
a suitable choice of hidden sector components.)

\subsection{Shift Embedding}

We consider a heterotic ${\mathbb Z}_6$-II orbifold which is embedded
into the gauge group E$_8 \times$E$_8$ as a shift with shift vector
\begin{equation}
V_6 = \left(
\frac{2}{3},0^5,\frac{1}{2},\frac{1}{2}\right)\left(\frac{1}{3},0^7\right) .
\end{equation}
Most of the time we will focus on the first E$_8$ only. The embedding
into the second E$_8$ is only important in the context of modular
invariance. The ${\mathbb Z}_3$ subgroup of ${\mathbb Z}_6$ is
embedded with a shift vector equivalent to
\begin{equation}
V_3 = 2V_6 = \left( \frac{4}{3}, 0^7\right) . 
\end{equation}
Within the first E$_8$, the roots 
\begin{equation}
\left( 0, \underline{ \pm 1, \pm 1, 0^5}\right)
\end{equation}
have an integer valued scalar product with $V_3$. Hence, the ${\mathbb
  Z}_3$ orbifold breaks the first E$_8$ to SO(14)$\times$U(1). We will
  see now that the embedding of the ${\mathbb Z}_2$ subgroup of
  ${\mathbb Z}_6$ yields SO(10) and continuous Wilson lines in ten
  dimensional representations. Indeed, the embedding of ${\mathbb
  Z}_2$ is specified by the shift vector
\begin{equation}
V_2 = 3V_6 = \left( 0^6, \frac{1}{2}, \frac{1}{2}\right) .
\end{equation}
Roots having integer valued scalar products with $V_3$ and $V_2$ (and
thus with $V_6$) are of the form
\begin{equation}\label{eq:so10}
\left( 0, \underline{\pm 1 , \pm 1, 0^3}, 0,0\right)
\hspace*{1in}
\mbox{SO(10) roots},
\end{equation}
\begin{equation}\label{eq:extrasu2s}
\left( 0^6, \pm 1, \pm 1\right) \hspace*{1in}
\mbox{SU(2)}\times\mbox{SU(2) roots} .
\end{equation}
The continuous Wilson lines which are of potential interest in the
context of electroweak symmetry breaking come from root operators
whose roots have half integer valued scalar product with $V_2$. That
applies to the following roots
\begin{equation} \label{eq:controots}
\left( 0 , \underline{\pm 1 , 0,0,0,0}, \underline{\pm 1, 0}\right) ,
\end{equation}
leading to a ({\bf 10},{\bf 2},{\bf 2}) representation of
SO(10)$\times$SU(2)$^2$. 

In order to hierarchically separate the GUT breaking from the
electroweak symmetry breaking we will project the SO(10) symmetry
further down
to the Standard Model symmetry at the compactification scale.
This can be done by introducing discrete Wilson lines. 
Along the lattice vectors of the SU(3) torus we switch on two
identical Wilson lines
\begin{equation}
a_3 = a_4 = \left( 0, \frac{2}{3}, -\frac{1}{3}, -\frac{1}{3},
0^4\right) 
\left( 0^8\right) .
\end{equation}
This Wilson line breaks our SO(10) group to
SU(3)$\times$SU(2)$\times$SU(2)$\times$U(1). In order, to obtain the
Standard Model group from SO(10) we need another discrete Wilson line
which corresponds to one of the SO(4) lattice vectors, e.g.\ $e_5$
\begin{equation}\label{eq:z2diswl}
a_5 = \left( \frac{1}{4},
\frac{1}{4},\frac{1}{4},\frac{1}{4},\frac{1}{4},-\frac{1}{4},
-\frac{1}{4}, \frac{1}{4}\right)\left( 0^6,\frac{1}{2},
\frac{1}{2}\right) .
\end{equation}
This finally, breaks SO(10) to
SU(3)$\times$SU(2)$\times$U(1)$\times$U(1), where the roots are given
by
 \begin{equation}\label{eq:su3xsu2}
\begin{array}{r c l}
\pm \left( 0^4 , 1, 1, 0, 0\right) & \rightarrow &
\mbox{SU(2)},\\
\pm \left( 0,\underline{1,-1,0}, 0^4\right) & \rightarrow &
\mbox{SU(3)}. 
\end{array}
\end{equation}
The decoupled U(1)s (with SO(10) embedding) are generated by
\begin{equation} \label{eq:u1in10}
H_5 - H_6 \,\,\, \mbox{and}\,\,\, H_2 + H_3 + H_4 .
\end{equation}
In particular the hypercharge corresponds to (see e.g.\
\cite{Zee:2003mt}) 
\begin{equation}
Y_W = \frac{2}{3}\left( H_2 + H_3 + H_4\right) -\left( H_5 -
H_6\right) .
\end{equation}

Now, let us have a closer look at continuous Wilson
lines. These will be associated to shifts with the SO(4)
lattice vector $e_6$.   
According to our general rules from section \ref{sec:wl}
we have to find a maximal set of mutually commuting hermitian operators built
from root operators corresponding to the roots in
(\ref{eq:controots}). These are given by 
\begin{equation}
E_{\gamma_i} + E_{-\gamma_i},\,\,\, i=1, \ldots ,4,
\end{equation}
with
\begin{equation}\label{eq:gammas}
\gamma_1 = \left( 0^4,1,0,1,0\right)  , \,\,\, \gamma_2 = \left(
0^4,1,0,-1,0\right)  , \,\,\, \gamma_3 = \left( 0^5, 1, 0
,1\right)  , \,\,\, \gamma_4 = \left( 0^5, 1, 0
,-1\right) .
\end{equation}
In the presence of discrete Wilson lines we can switch on only
continuous Wilson lines which commute with the discrete
ones. This leaves only the operators corresponding to $\gamma_1$ and
$\gamma_3$. Consider for instance
\begin{equation}\label{eq:conwilemz6}
T = \lambda \left( E_{\gamma_1}+ E_{-\gamma_1}\right) .
\end{equation}
This Wilson line breaks the SU(2) in (\ref{eq:su3xsu2}) times the U(1)
generated by $H_5 - H_6$ (see (\ref{eq:u1in10})) to a U(1) with
generator $H_6$. The SU(3) and the other $U(1)$ in (\ref{eq:u1in10})
commute with the continuous Wilson line. Identifying the weak isospin
as the U(1) generator of the broken SU(2) 
\begin{equation}
I_W ^3 = \frac{1}{2} \left( H_5 + H_6 \right)
\end{equation}
we find that the electromagnetic U(1)
\begin{equation}
Q_{em} = I^3 _W + Y_W /2 = \frac{1}{3} \left( H_2 + H_3 + H_4\right) +
H_6 
\end{equation}
is left unbroken. In addition, there will be another unbroken U(1) which is
embedded into SO(10)
\begin{equation}
H_2 + H_3 + H_4 - H_6.
\end{equation}
That this extra U(1) appears is a consequence of the construction
breaking the SO(10) by discrete Wilson lines to the Standard Model
group. Since discrete Wilson lines do not reduce the rank there will be
always one additional U(1) at that stage. The rank reducing property of our
Wilson line (\ref{eq:conwilemz6}) was used to lower the rank of
isospin times hypercharge by one. 

So far, we have focused on the breaking pattern within the
intermediate SO(10). Here, the considered continuous Wilson line has
indeed the features of a standard model Higgs (or more precisely the
MSSM pair, $E_{\gamma_1}$ and $E_{-\gamma_1}$ carry opposite
hypercharges). If we come back to the string model we started with,
there 
are some caveats to be mentioned. Before we switched on the discrete
Wilson lines breaking SO(10) to the SM group there were in addition
to SO(10) also two unbroken SU(2) groups
(\ref{eq:extrasu2s}). One of these SU(2)s is
broken by the discrete Wilson lines and the other one by the
continuous Wilson line. 
The picture can look different in other models
as will be seen in the next section.

\section{A ${\mathbb Z}_2 \times {\mathbb Z}_2$ Example}

\begin{figure}[h!]
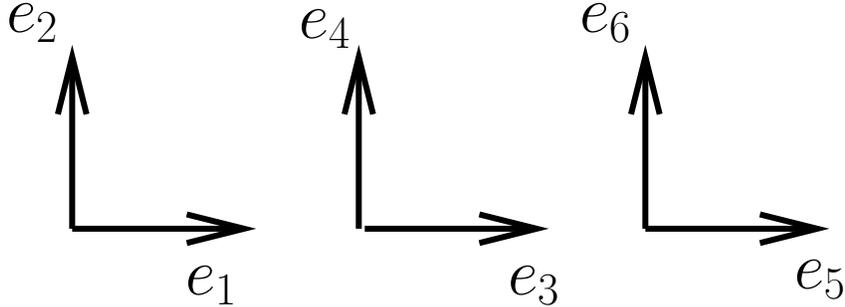

\centering
\begin{center}\input compspace.pstex_t \end{center}
\caption{The six dimensional compact space.}
\label{fig:compspace}
\end{figure}

As the compact space we take the product of three SO(4) lattices
(figure \ref{fig:compspace}) modded by a ${\mathbb Z}_2 \times
{\mathbb Z}_2$ symmetry.
The generators of ${\mathbb Z}_2 \times
{\mathbb Z}_2$ act as rotations defined by the vectors (see
discussion around (\ref{eq:z6IIspace}) for an explanation of the notation)
\begin{equation}
v_1 = \left( \frac{1}{2}, -\frac{1}{2}, 0\right) \,\,\, , \,\,\, v_2 =
\left( 0, \frac{1}{2}, -\frac{1}{2}\right) .
\end{equation}
For more details of the geometry and the fixed point structure see
e.g.\  \cite{Forste:2004ie}.

Before going into details of the model we first outline the rough
picture. 
The embedding of the orbifold into the gauge group E$_8\times$E$_8$
will be such that 
one of the ${\mathbb Z}_2$ breaks the visible E$_8$ to
a subgroup containing SO(12). The other generator breaks the SO(12) to
SO(10). This falls into the general pattern discussed in section
\ref{sec:ewstrat} and thus gives rise to continuous Wilson lines in
ten dimensional representations of SO(10). Discrete Wilson lines will
break the SO(10) down to the Standard Model group with an extra U(1)
and reduce the continuous Wilson lines to SU(2) doublets.

\subsection{Breaking E$_8$ to SM via SO(10)}

Now, we provide the details of the construction. Again, we consider
the hidden E$_8$ only when it is important for satisfying modular
invariance conditions.
The ${\mathbb Z}_2$
generated by $v_1$ is shift embedded into E$_8 \times$E$_8$ with 
\begin{equation}
V_1 = \left( \frac{1}{2}, -\frac{1}{2}, 0^6\right)\left( 1, 0^7\right)
, 
\end{equation}
breaking E$_8\times$E$_8 ^\prime$ to E$_7
\times$SU(2)$\times$SO(16)$^\prime$. 
The orbifold twist acts as 180$^{\mbox{\scriptsize o}}$ rotations in the
first two tori in figure 
\ref{fig:compspace}. Within these two tori we also introduce a
discrete Wilson line which we associate with shifts along
$e_1$
\begin{equation}
 a_1  =  \left( 1, 0^7\right)\left( 0^2, \frac{1}{2},\frac{1}{2},0^2,
 \frac{1}{2}, -\frac{1}{2} 
\right) .
\end{equation}
The last torus (spanned by $e_5$ and $e_6$) is invariant under the
actions of $v_1$ and $a_1$. The roots of operators belonging to the
gauge symmetry in the bulk of the corresponding orbifold have integer valued
scalar products with $V_1$ and $a_1$. Within the first E$_8$ this
applies to
\begin{equation}\label{eq:bulkgroup}
\begin{array}{r c l}
\left( \pm 1 , \pm 1, 0^6\right) & \rightarrow & \mbox{SU(2)}^2 , \\
\left( 0, 0, \underline{ \pm 1 , \pm 1 ,0 ,0 ,0 ,0}\right) &
\rightarrow & \mbox{SO(12)}.
\end{array} 
\end{equation}
Hence the gauge group in the bulk of the third lattice in figure
\ref{fig:compspace} is
\begin{equation}
\mbox{SO(12)}\times \mbox{SU(2)} \times \mbox{SU(2)}.
\end{equation}
The next step is to break that gauge group to an SO(10) intermediate
group. This can be achieved by embedding the ${\mathbb Z}_2 \times
{\mathbb Z}_2$ generator $v_2$ with a shift
\begin{equation}
 V_2 = \left( 0, \frac{1}{2}, -\frac{1}{2}, 0^5\right)
\left( 0^8\right) .
\end{equation}
Indeed, SO(12)$\times$SU(2)$^2$ roots (\ref{eq:bulkgroup}) having
integer valued scalar product with $V_2$ are
\begin{equation}\label{eq:intermso10}
\begin{array}{l c r}
\left( 0^3 , \underline{\pm 1 , \pm 1, 0,0,0}\right) & \rightarrow &
\mbox{SO(10)} \end{array} .
\end{equation}
In contrast to our previous example the extra SU(2) factors in
(\ref{eq:bulkgroup}) are now broken at the compactification scale. The
continuous Wilson lines which are of potential interest for
electroweak symmetry breaking correspond to SO(12) roots having half
integer scalar product with $V_2$
\begin{equation}\label{eq:pothiggs}
\left(0,0,\pm1,\underline{\pm 1,0,0,0,0}\right) .
\end{equation}
These form two {\bf 10} representations of SO(10).

\begin{figure}[h!]
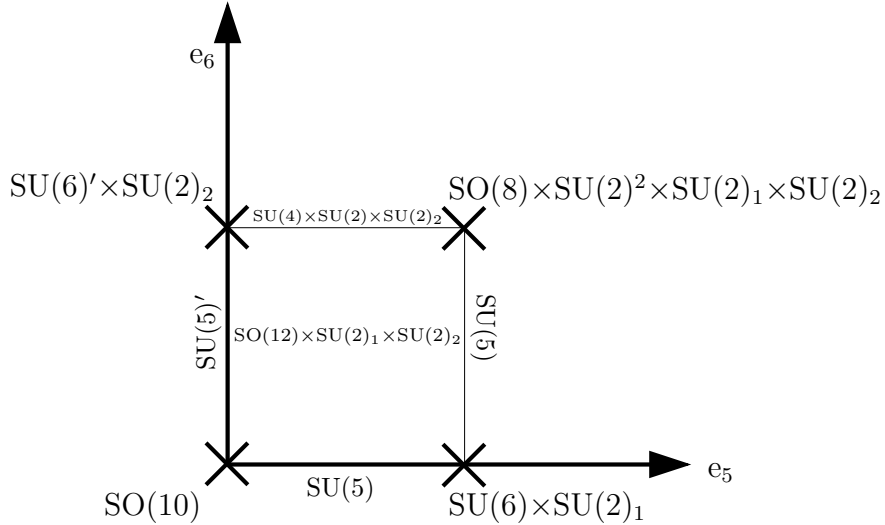

\centering
\begin{center}\input ztwo-geog.pstex_t \end{center}
\caption{Local projections in the third torus. U(1) factors are
  suppressed. A prime indicates a 
  different embedding into the bulk group. Groups written at lines
  connecting two fixed points correspond to simultaneously imposing
  the projection conditions of both fixed points. The gauge symmetry
  in four dimensions is SU(3)$\times$SU(2). }
\label{fig:ztwo-geog}
\end{figure}

In the following we will break SO(10) further down to the Standard
Model group by means of discrete Wilson lines. We will discuss two
setups with different gauge group geographies in the extra dimensions
and identical pictures in four dimensions. 
In the first model we introduce two discrete Wilson lines within the
third torus
\begin{equation}\label{eq:disWLin3rdtorus}
\begin{array}{r c l}
a_5  & = & \left( 
\frac{1}{4}, \frac{1}{4}, \frac{1}{4}, \frac{1}{4}, \frac{1}{4},
\frac{1}{4}, \frac{1}{4}, \frac{1}{4}\right) \left( 0, \frac{1}{2},
-\frac{1}{2}, 0^5\right) , \\ a_6 & = & \left( -\frac{1}{4},
\frac{1}{4}, \frac{1}{4},- \frac{1}{4},- \frac{1}{4},- \frac{1}{4},
\frac{1}{4}, \frac{1}{4}\right)\left( \frac{1}{4}, \frac{1}{4},
\frac{1}{4}, \frac{1}{4}, \frac{1}{4}, \frac{1}{4}, \frac{1}{4},
\frac{1}{4}\right) .  
\end{array} 
\end{equation}
SO(10) roots having integer valued scalar products with these two
Wilson lines are
\begin{equation}\label{eq:smgroup}
\begin{array}{r c l}
\left( 0^3, \underline{1,-1,0}, 0,0\right) & \rightarrow &
\mbox{SU(3)} , \\
\left( 0^6,\underline{1,-1}\right) & \rightarrow & \mbox{SU(2)} ,
\end{array} 
\end{equation}
and hence we succeeded to obtain a four dimensional Standard Model
group which is embedded into an intermediate SO(10). Among the additional
U(1) factors one combination corresponds to the hypercharge 
\begin{equation}\label{eq:hypergen}
Y_W = \frac{2}{3}\left( H_4 + H_5 + H_6\right) - H_7 - H_8 .
\end{equation}
Zooming in on the third torus in figure \ref{fig:compspace} we obtain
the discussed breaking in figure \ref{fig:ztwo-geog}. 

\begin{figure}[h!]
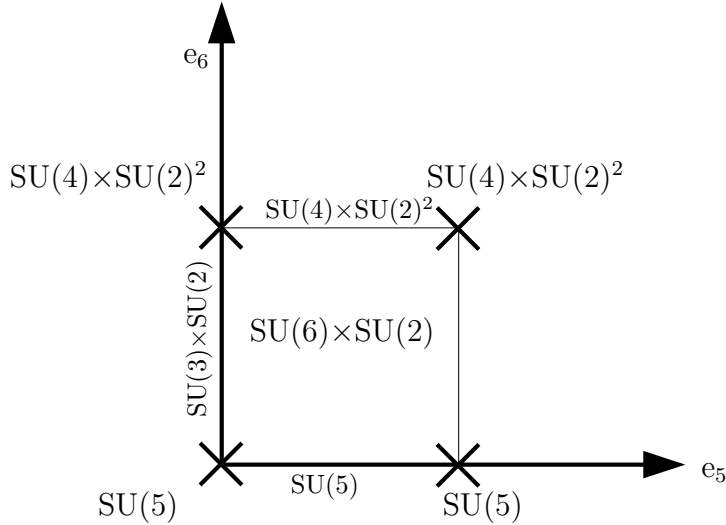

\centering
\begin{center}\input geog-ztwo.pstex_t \end{center}
\caption{Figure \ref{fig:ztwo-geog} changes into the present figure if
  the Wilson line $a_5$ is moved to the direction $e_3$. The spectrum
  at fixed points separated along the horizontal direction is identical.}
\label{fig:geog-ztwo}
\end{figure}

The second setup we want to consider is closely related to the one
discussed so far. We merely associate the Wilson line $a_5$ to the cycle
generated by $e_3$
\begin{equation}
a_5 \to a_3.
\end{equation}
The only change is a modification of figure \ref{fig:ztwo-geog} since
now the Wilson line $a_3$ breaks the bulk group in the depicted torus.
The modified picture 
is shown in figure \ref{fig:geog-ztwo}.

\subsection{Electroweak Breaking}

\begin{figure}[h!]
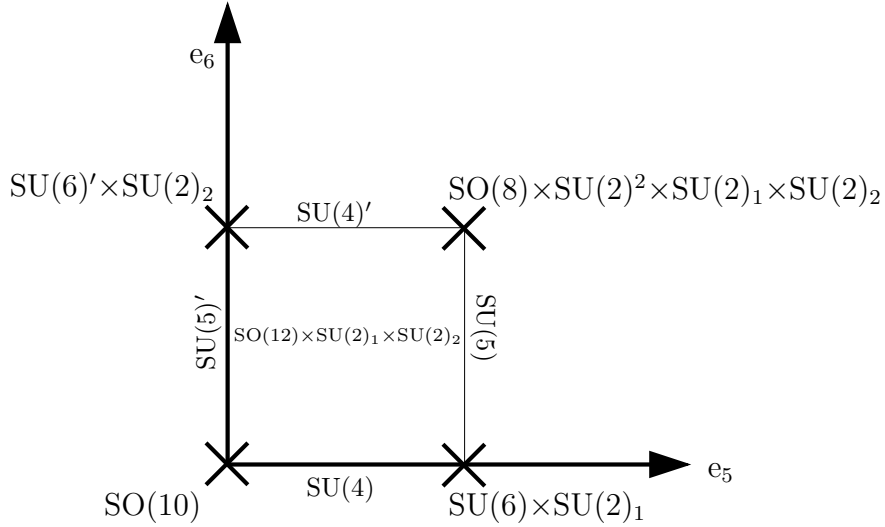

\centering
\begin{center}\input broken.pstex_t \end{center}
\caption{Compared to figure \ref{fig:ztwo-geog} a continuous Wilson
  line is added in the $e_5$ direction. The unbroken gauge group in
  four dimensions is SU(3)$\times$U(1)$^5$. }
\label{fig:broken}
\end{figure}

Finally, we turn on a continuous Wilson line. This will be associated
with lattice shifts by $e_5$, i.e.\ we superpose the discrete Wilson
line $a_5$ (\ref{eq:disWLin3rdtorus}) with a continuous Wilson
line. From root operators belonging to the set (\ref{eq:pothiggs}) we
form a maximal set of hermitian and mutually commuting operators. In
addition, these operators should commute with all discrete Wilson
lines. In other words, we consider only a subset of roots in
(\ref{eq:pothiggs}) which have vanishing scalar products with the
discrete Wilson lines. For a generic continuous Wilson line one finds
\begin{equation}\label{z2texcontW}
T = \lambda_1 C_1 + \lambda_2 C_2 .
\end{equation}
with
\begin{equation}
C_i = E_{\gamma_i} + E_{-\gamma_i}\,\,\, i = 1,2.
\end{equation}
and
\begin{equation}
\gamma_1 = \left( 0^2,1,0^3,-1,0\right) \,\,\, , \,\,\, \gamma_2 =
\left( 0^2, 1, 0^4, -1\right) .
\end{equation}
As an example we consider the case $\lambda_2 = 0$. The following
seven Cartan generators of E$_8$ commute with the continuous Wilson
line
\begin{equation}
H_1,\,\,\, H_2,\,\,\, H_3 + H_7,\,\,\, H_4,\,\,\, H_5,\,\,\,
H_6,\,\,\, H_8.
\end{equation}
The last five generators are embedded into the intermediate SO(10)
(\ref{eq:intermso10}). The combinations $H_4 -H_5$ and $H_5 - H_6$
form the Cartan algebra of the SU(3) colour which is unbroken also
after the continuous Wilson line has been turned on (see
(\ref{eq:smgroup})). The generator $H_7 - H_8$ is the Cartan generator
of the SU(2) in (\ref{eq:smgroup}) and hence its charges correspond to
the weak isospin. This SU(2) is completely broken by the continuous
Wilson line. The hypercharge generator (\ref{eq:hypergen})
does not commute with $T$, either. However, the combination of isospin
and hypercharge 
generating the electromagnetic U(1) is left unbroken
\begin{equation}
Q_{em} = I_W^3 + Y_W/2 = \frac{1}{2}\left( H_7 - H_8\right) +Y_W/2 =
\frac{1}{3} \left( H_4 + H_5 + H_6\right) - H_8 .
\end{equation}
Thus our continuous Wilson line breaking produces exactly the
electroweak breaking of the MSSM. Focusing on the third torus in
figure \ref{fig:compspace} we draw a geometric picture of the
electroweak breaking in figure  \ref{fig:broken}. 

\begin{figure}[h!]
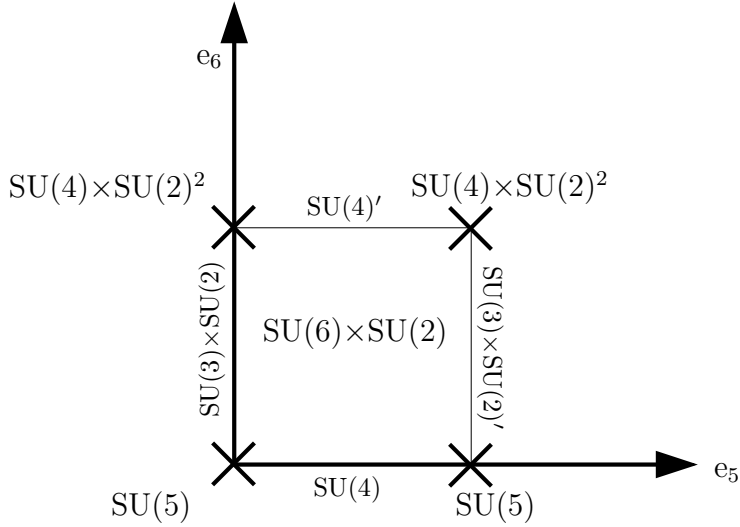

\centering
\begin{center}\input brokenalt.pstex_t \end{center}
\caption{The continuous Wilson line lifts the degeneracy of
  horizontally separated fixed points.}
\label{fig:brokenalt}
\end{figure}

In the alternative setup of figure \ref{fig:geog-ztwo} we do not need
to superpose a continuous with a discrete Wilson line. From our
previous discussion only the geometric picture drawn in figure
\ref{fig:broken} is altered into the picture drawn in figure
\ref{fig:brokenalt}. 

In summary, we have achieved some improvements compared to the model
discussed in section \ref{eq:z6ex}. Within the visible E$_8$ all non
Abelian gauge symmetries are broken to the Standard Model symmetry at
the compactification scale. The continuous Wilson line just breaks
\begin{equation}
\mbox{SU(2)}_W \times \mbox{U(1)}_Y \to \mbox{U(1)}_{em} .
\end{equation}
%

\section{Conclusions}

In the present paper we investigated symmetry breaking due to
continuous Wilson lines. Previously the mechanism of continuous Wilson
line breaking had been introduced in models where the orbifold is
embedded as a rotation. We were able to rephrase that technique such
that it works also in shift embedded orbifolds. The geometric picture
of \cite{Forste:2005rs} is confirmed in the new framework. Like a
discrete Wilson line the continuous Wilson line also lifts the
degeneracy between fixed points. The unbroken gauge symmetry, however,
is the same at each fixed point. The continuous Wilson line is responsible
for a misalignment of the embedding into the bulk group. 

As a first application we studied the model of \cite{Forste:2005rs}
reformulated as a shift embedded orbifold. Results of
\cite{Forste:2005rs} are confirmed and moreover the splitting of bulk
matter is given. The effective four dimensional picture corresponds to
the Higgs mechanism where matter becomes massive due to Yukawa
couplings to the Higgs field. For bulk matter we did not have to
compute these Yukawa couplings explicitly since they are fixed by higher
dimensional gauge invariance.  

In our models with unbroken $N=1$ supersymmetry the scale for symmetry
breaking by continuous Wilson lines is associated to a flat direction
in moduli space \cite{Font:1988tp,Font:1988mm}. It can be anywhere
between zero and the compactification scale. (In the
decompactification limit higher dimensional gauge symmetry is restored
and the continuous Wilson line becomes a higher dimensional gauge
degree of freedom.) 
In realistic models with broken supersymmetry that flat direction
should be lifted and the breaking scale will be fixed. We leave the
investigation of such a mechanism for future work. In the present
paper, we studied the 
possibility that the breaking scale is given by the electroweak scale
as far as the gauge symmetry breaking patterns are concerned. In a
class of models where the Standard Model group appears as a subgroup
of some intermediate SO(10) we developed a strategy how to obtain
electroweak symmetry breaking from continuous Wilson line
breaking. The intermediate SO(10) originates from the breaking
of a larger group. This is typically the case in heterotic
orbifolds. Under the orbifold breaking to SO(10) the adjoint of the
higher dimensional group should branch into the adjoint of SO(10) and
ten dimensional representations. In the subsequent breaking a Higgs
doublet can be obtained from the {\bf 10} of SO(10). Such a Higgs
doublet corresponds to a continuous Wilson line. 

As examples we studied a ${\mathbb Z}_6$-II and a ${\mathbb Z}_2
\times {\mathbb Z}_2$ heterotic orbifold. For the ${\mathbb Z}_6$-II
prescriptions in rotational as well as shift embedding are given. The
connection between the two approaches is further illustrated. Although
the electroweak symmetry breaking can be achieved by a continuous
Wilson line this model has some shortcomings. There are additional
non Abelian symmetries which are also broken only by the continuous
Wilson line. The situation improves for the   ${\mathbb Z}_2
\times {\mathbb Z}_2$ model. Apart form the Standard Model group all
non Abelian groups are broken at the compactification scale by
orbifold and discrete Wilson lines. A continuous Wilson line providing
exactly electroweak symmetry breaking has been identified. 

Many questions are left open for future research. In our models for
electroweak symmetry breaking we focused on the gauge sector only. A
realistic model should provide also the matter spectrum with the
correct quark and lepton masses. The geometric understanding of
spontaneous symmetry breaking, discussed in our paper, can serve
the intuition in the search for realistic models.

\bigskip

\noindent {\bf Acknowledgements}

\noindent
This work was
partially supported by the European community's 6th framework
programs MRTN-CT-2004-503369 ``Quest for Unification'' and
MRTN-CT-2004-005104 ``Forces Universe''. We would like to thank
Patrick Vaudrevange for useful discussions.

\begin{appendix}

\section{Rotational versus Shift Embedding in the ${\mathbb Z}_3$
  Case}\label{ap:z3} 

In the following, we will discuss how our considerations in section 2
are modified if we study a ${\mathbb Z}_3$ orbifold instead of
${\mathbb Z}_2$. For simplicity, we consider just two extra
dimension and take for the bulk group SU(3). The extra dimensions are
compactified on an SU(3) root lattice. First we will embed the
${\mathbb Z}_3$ as a rotation into SU(3) and discuss the continuous
Wilson lines. This will be akin to the discussion of
\cite{Ibanez:1987xa} where the same embedding is given for SU(3)
subgroups of E$_8\times$E$_8$. Afterwards we will consider these continuous
Wilson lines in shift embeddings.

\subsection{Rotational Embedding}

The discussion here will be on a schematic level and we will not
construct the full algebra automorphism corresponding to the embedding
of ${\mathbb Z}_3$. The six roots of SU(3) are
\begin{equation}
\alpha_1 = \left( \sqrt{2}, 0\right)\,\,\, , \,\,\, \alpha_2 = \left(
-\frac{\sqrt{2}}{2}, \frac{\sqrt{6}}{2}\right)\,\,\, , \,\,\,
-\alpha_1\,\,\, , \,\,\, -\alpha_2 \,\,\, , \,\,\, \pm\left( \alpha_1
+ \alpha_2\right) .
\end{equation}
We specify the rotational embedding of ${\mathbb Z}_3$ by the rotation
matrix
\begin{equation} \label{eq:z3rot}
\Theta = \left( \begin{array}{c c} -\frac{1}{2} &
  -\frac{\sqrt{3}}{2}\\
\frac{\sqrt{3}}{2} & -\frac{1}{2} \end{array}\right) ,
\end{equation}
acting on the vector of Cartan generators $H = \left( H_1 , H_2 \right)^T$.
The embedding appears as a cyclic permutation within two subsets
of the roots consisting of three roots each
\begin{equation}
\alpha_1 \stackrel{\Theta}{\longrightarrow} \alpha_2
\stackrel{\Theta}{\longrightarrow} -\alpha_1 - \alpha_2 \,\,\, ,
\,\,\,
-\alpha_1 \stackrel{\Theta}{\longrightarrow} -\alpha_2
\stackrel{\Theta}{\longrightarrow} \alpha_1 + \alpha_2 .
\end{equation}
The eigenvalues of $\Theta$ are $\exp \left( \pm 2\pi i /3\right)$ and
hence there are no invariant directions. The Cartan generators $H_1$
and $H_2$ are projected out. There are two invariant combinations of
root generators (suppressing possible phase factors appearing when
lifting the embedding to an algebra automorphism)
\begin{equation}
E_{\alpha_1} + E_{\alpha_2} + E_{-\alpha_1 - \alpha_2} \,\,\, , \,\,\,
E_{-\alpha_1} + E_{-\alpha_2} + E_{\alpha_1 + \alpha_2}.
\end{equation}
From these we can construct two commuting hermitian operators which
serve as new Cartan generators
\begin{eqnarray}
H^{new}_1 & = & E_{\alpha_1} + E_{\alpha_2} + E_{-\alpha_1 - \alpha_2} + 
E_{-\alpha_1} + E_{-\alpha_2} + E_{\alpha_1 + \alpha_2} , \\
H^{new}_2 & = & i\left( E_{\alpha_1} + E_{\alpha_2} + E_{-\alpha_1 -
  \alpha_2} -  
E_{-\alpha_1} - E_{-\alpha_2} - E_{\alpha_1 + \alpha_2}\right) .
\end{eqnarray}
 The unbroken gauge group is
$$\mbox{ U(1)}^2 . $$
As explained in \cite{Ibanez:1987xa} the continuous Wilson line
can be parameterised
along SU(3) weight directions
\begin{equation}
W = \lambda_1 w_1 \cdot H
+ \lambda_2 w_2 \cdot H,
\end{equation}
where $w_i$ are the fundamental weights which are dual to the lattice
vectors $\alpha_i$
\begin{equation}
w_1 = \left( \frac{1}{\sqrt{2}}, \frac{1}{\sqrt{6}}\right)
 \,\,\, ,
\,\,\, w_2 = \left( 0, \frac{2}{\sqrt{6}}\right) .
\end{equation}
The two cycles of the compactification torus are identified by the
${\mathbb Z}_3$ orbifold action. Therefore on the other cycle the
Wilson line has to be the ${\mathbb Z}_3$ image
\begin{equation}
\Theta W =   \lambda_1\left( w_2 -w_1 \right) \cdot H
- \lambda_2 w_1  \cdot H,
\end{equation}
Already if only one of the $\lambda_i$ is neither zero nor an integer
SU(3) is completely broken in the presence of a continuous Wilson
line.

Before reproducing this result in shift embedding let us also consider
the geometric picture for the symmetry breaking due to continuous
Wilson lines. To this end, we first specify the fixed points within
the compact space. In order to avoid confusion with the gauge group we
choose non normalised basis vectors for the compactification space
\begin{equation}
e_1 = R \alpha_1 \,\,\, , \,\,\, e_2 = R\alpha_2 ,
\end{equation}
where $\alpha_i$ are the SU(3) root vectors introduced earlier.
The ${\mathbb Z}_3$ action on the compactification lattice is given by
the same $\Theta$ as in (\ref{eq:z3rot}).
There are three ${\mathbb Z}_3$ fixed points on the torus
\begin{equation}
P_1 = \mbox{O} \,\,\, , \,\,\, P_2 = \frac{2R}{3} \alpha_1 +
\frac{R}{3} \alpha_2 \,\,\, , \,\,\, P_3 = \frac{R}{3} \alpha_1 +
\frac{2R}{3} \alpha_2 .
\end{equation}
The point $P_1$ is a fixed point in the complex plane, whereas the
${\mathbb Z}_3$ image of $P_2$ ($P_3$) has to be shifted by
$R\alpha_1$ ($R\alpha_1 + R \alpha_2$) in order to obtain the
preimage. This means that the projections are with respect to $\Theta$
at $P_1$ w.r.t.\ $\Theta$ times the adjoint action of $\exp\left[
  2\pi i W\right]$ at $P_2$ and w.r.t.\ $\Theta$ times the adjoint action
of $\exp\left[ -2 \pi i \Theta^2 W\right]$ at $P_3$. For simplicity
let us consider the case $\lambda_2 = 0$. The algebra
elements invariant under the local projections are ($\Theta^2 w_1 =
-w_2$)
\begin{equation}
\begin{array}{l c r}
\begin{array}{l}
E_{\alpha_1} + E_{\alpha_2} + E_{-\alpha_1 - \alpha_2} + 
\mbox{h.c.} , \\
i\left( E_{\alpha_1} + E_{\alpha_2} + E_{-\alpha_1 -
  \alpha_2} -  
\mbox{h.c.}\right) 
\end{array} & \mbox{at} & P_1 , \\[3ex]
\begin{array}{l}
E_{\alpha_1} + E_{\alpha_2} + e^{-2\pi i \lambda_1 }E_{-\alpha_1 -
  \alpha_2} + \mbox{h.c.}  , \\
i\left( E_{\alpha_1} + E_{\alpha_2} + e^{-2\pi i \lambda_1 }E_{-\alpha_1 -
  \alpha_2} -  \mbox{h.c.}\right) 
\end{array} & \mbox{at} & P_2 , \\[3ex]
\begin{array}{l}
E_{\alpha_1} + e^{2\pi i\lambda_1} E_{\alpha_2} + E_{-\alpha_1 - 
  \alpha_2} + \mbox{h.c.}  , \\
i\left( E_{\alpha_1} + e^{2 \pi i \lambda_1} E_{\alpha_2} + E_{-\alpha_1 -
  \alpha_2} -  \mbox{h.c.}\right) 
\end{array} & \mbox{at} & P_3.

\end{array}
\end{equation}
At each of the fixed points there is a U(1)$^2$ gauge symmetry. For
generic $\lambda_1$ these U(1)s are all embedded differently into the
bulk SU(3) and the gauge group is broken completely.

\subsection{Shift Embedding}

As in the ${\mathbb Z}_2$ case we use the space group notation. The
orbifold is taken to be shift embedded
\begin{equation}
\left( \theta , 0\right):\,\,\, \begin{array}{l l l l}
H_i & \to & H_i , & i = 1, \ldots, r \\
E_{\alpha_k} & \to & \mbox{exp}\{ 2 \pi i \alpha_k \cdot V\}
E_{\alpha_k}, & k 
  = 1, \ldots , \mbox{dim}(G) - r \end{array} ,
\end{equation}
i.e.\ we embed the orbifold as the adjoint action with
$$ e^{2\pi i V^i H_i}. $$

For the projections at the other two fixed points we also need
\begin{equation} \label{eq:wlgenz3}
\left( \theta , R\alpha_1 \right):\,\,\, \begin{array}{l l l l}
H_i & \to & t H_i t^{-1} , & i = 1, \ldots, r \\
E_{\alpha_k} & \to & \mbox{exp}\{ 2 \pi i \alpha_k \cdot V\}
t E_{\alpha_k}\, t^{-1}, & k 
  = 1, \ldots , \mbox{dim}(G) - r \end{array} ,
\end{equation}
and
\begin{equation} \label{eq:wlgenz3-2}
\left( \theta , R\alpha_1 + R\alpha_2 \right):\,\,\, \begin{array}{l l l l}
H_i & \to & t_1 H_i t_1 ^{-1} , & i = 1, \ldots, r \\
E_{\alpha_k} & \to & \mbox{exp}\{ 2 \pi i \alpha_k \cdot V\}
t_1 E_{\alpha_k}\, t_1 ^{-1}, & k 
  = 1, \ldots , \mbox{dim}(G) - r \end{array} .
\end{equation}
Again, we describe a Wilson line by some general adjoint
transformation with constant group elements $t, t_1$.
From the orbifold identification $-R\left(\alpha_1 + \alpha_2\right) =
\theta^2 R\alpha_1$ one obtains the condition
\begin{equation}\label{eq:secwl}
t_1 = e^{4\pi i V^i H_i} t^{-1}  e^{-4\pi i V^i H_i} .
\end{equation}
The condition that the space group element leaving the point $P_2$
invariant is a ${\mathbb Z}_3$ action yields the consistency condition
\begin{equation}\label{eq:consistency}
t\,  \left( e^{2\pi i V^i H_i}\, t\,  e^{-2\pi i V^i H_i}\right)\, \left(
e^{4\pi i V^i H_i}\, t\, 
e^{-4\pi i V^i H_i}\right) = 1 ,
\end{equation}
which should also hold with arbitrary permutations of the three
factors (${\mathbb Z}_3$ is Abelian). Then, the relation (\ref{eq:secwl})
implies that the embedding of the element leaving $P_3$ fixed is
consistent. In addition, $t$ should be unitary in order to ensure that
the orbifold maps hermitian operators to hermitian operators.

Now, let us discuss solutions to (\ref{eq:consistency}). If $t$ is an
element of the Cartan subgroup the condition implies $t^3 =1$. This
corresponds to a discrete Wilson line and will not be discussed
further here. For the solution corresponding to a continuous Wilson
line we make the ansatz
\begin{equation}
t= e^{2 \pi i \left( T + T^\dagger\right)}
\end{equation}
ensuring unitarity of $t$. Further we choose $T$ to be an eigenstate
of the orbifold action, whose eigenvalue we take without loss of
generality to 
be given by
\begin{equation}
e^{2\pi i\, V^i H_i} \, T\, e^{-2\pi i\, V^i H_i} = e^{\frac{2 \pi i}{3}} T .
\end{equation}
In difference to the ${\mathbb Z}_2$ case the eigenvalue is not real
and $T^\dagger$ transforms with a different phase. Therefore, the
${\mathbb Z}_3$ case differs from the ${\mathbb Z}_2$ case since now
we also need to impose 
\begin{equation}
\left[ T, T^\dagger \right] = 0.
\end{equation}
In particular it will be impossible to equate $T$ with a single root
operator, now. This will be important when we explicitly connect this
construction to the earlier discussed rotational embedding. If all the
above conditions are satisfied (\ref{eq:consistency}) is solved thanks
to the identity
$$ 1 + e^{\frac{2\pi i}{3}} + e^{\frac{ 4 \pi i}{3}} = 0 .$$

As in the ${\mathbb Z}_2$ case we can write the action of the space
group element in (\ref{eq:wlgenz3}) as a shift embedding w.r.t.\ to a
conjugated Cartan Weyl basis
\begin{eqnarray}
\hat{H}_i &  = & t^{2/3}\left( e^{2\pi i\, V^i H_i} \, t\, e^{-2\pi
  i\, V^i H_i}\right) ^{1/3}\,  H_i \, \left( e^{2\pi i\, V^i H_i} \, t\,
  e^{-2\pi 
  i\, V^i H_i}\right) ^{-1/3}t^{-2/3}, \\
\hat{E}_{\alpha} & = & \, t^{2/3}\left( e^{2\pi i\, V^i H_i} \, t\, e^{-2\pi
  i\, V^i H_i}\right) ^{1/3} \, E_\alpha \, \left( e^{2\pi i\, V^i H_i} \, t\,
  e^{-2\pi 
  i\, V^i H_i}\right) ^{-1/3}t^{-2/3}.
\end{eqnarray}
Similar considerations apply to the fixed point $P_3$. This reproduces
the geometrical picture that the unbroken gauge group at all fixed
points is the same, but the embedding into the bulk group differs from
fixed point to fixed point.

Let us now come back to our particular SU(3) example from the previous
section. We choose the shift vector
\begin{equation}
V = \frac{1}{3} \alpha_1 .
\end{equation}
This implies that under the orbifold the Cartan generators are
invariant, $E_{\alpha_1}$, $E_{\alpha_2}$ and $E_{-\alpha_1 -
  \alpha_2}$ transform with a phase $\exp\left\{ - 2\pi i /3\right\}$
and the remaining root generators transform with the complex
conjugated phase. Thus, none of the root generators is invariant under
the orbifold and the unbroken gauge group is
\begin{equation}
\mbox{U(1)}^2 .
\end{equation}

Now, let us turn on a continuous Wilson line. As in the ${\mathbb
  Z}_2$ case wee seek a maximal set of hermitian mutually commuting
  generators which transform non trivially under the orbifold. In
  addition, terms within those generators transforming differently
  under the orbifold have to commute. In our case, such a set is given
  by
\begin{eqnarray}
C_1 & = & E_{\alpha_1} + E_{\alpha_2} + E_{-\alpha_1 - \alpha_2} + 
E_{-\alpha_1} + E_{-\alpha_2} + E_{\alpha_1 + \alpha_2} , \\
C_2 & = & i\left( E_{\alpha_1} + E_{\alpha_2} + E_{-\alpha_1 -
  \alpha_2} -  
E_{-\alpha_1} - E_{-\alpha_2} - E_{\alpha_1 + \alpha_2}\right) ,
\end{eqnarray}
and the continuous Wilson line can be parameterised e.g.\ as
\begin{equation}
T + T^\dagger = \lambda_1 C_1 + \lambda_2 C_2 .
\end{equation}
Neither $C_1$ nor $C_2$ commute with any combination of the Cartan
generators and hence the gauge symmetry is broken completely as long
as at least one of the $\lambda_i$ is non vanishing. 

Let us discuss the connection to rotational embeddings, now. Due to
their defining properties $C_1$ and $C_2$ can serve as an alternative
set of Cartan generators. Under the originally shift embedded orbifold
they transform as
\begin{equation}
\left(
\begin{array}{c} 
C_1 \\ C_2 \end{array}\right) \longrightarrow
\Theta\left(\begin{array}{c}  
C_1 \\ C_2 \end{array}\right) ,
\end{equation}
where $\Theta$ is the same rotation matrix as in
(\ref{eq:z3rot}). 
Thus w.r.t.\ that set of Cartan generators the shift embedded orbifold
looks the same as the rotationally embedded orbifold considered in the
previous section and the description of continuous Wilson lines agrees
with the previous one. 

The presentation of ${\mathbb Z}_2$ and  ${\mathbb Z}_3$ should
suffice to render the discussion of continuous Wilson lines in any
${\mathbb Z}_N$ orbifold straightforward.

\section{Invariant Combinations of Root Generators}\label{ap:e6}

In this appendix we derive equation (\ref{eq:invcomb}). By using formul\ae\ 
as given in appendix A of \cite{Forste:2005rs} one can compute the
transformed $E_{\alpha}$ and $E_{\beta}$
\begin{eqnarray}
tE_\alpha t^{-1} & = & \cos \left( \lambda_2 N_{\gamma, \alpha}\right)
  \cos\left( \lambda_1 N_{\delta,\alpha}\right)E_{\alpha} + i
  \sin\left( \lambda_2 N_{\gamma, \alpha}\right) \cos\left( \lambda_1
  N_{\delta, \alpha}\right) E_{\gamma + \alpha}\nonumber \\ & & + i
  \cos\left( 
  \lambda_2 N_{\gamma, \alpha + \delta}\right) \sin\left( \lambda_1
  N_{\delta,\alpha} \right) E_{\alpha + \delta} - \sin\left( \lambda_2
  N_{\gamma, \alpha + \delta}\right) \sin\left( \lambda_1
  N_{\delta,\gamma}\right) E_\beta ,\label{eq:atrans}\\
t E_\beta t^{-1} & = &  \cos \left( \lambda_2 N_{-\gamma, \beta}\right)
  \cos\left( \lambda_1 N_{-\delta,\beta}\right)E_{\beta} + i
  \sin\left( \lambda_2 N_{-\gamma, \beta}\right) \cos\left( \lambda_1
  N_{-\delta, \beta}\right) E_{\alpha + \delta}\nonumber \\ & &
  \hspace*{-.3in} 
+ i\cos\left( 
  \lambda_2 N_{-\gamma, \beta - \delta}\right) \sin\left( \lambda_1
  N_{-\delta,\beta} \right) E_{\alpha + \gamma} - \sin\left( \lambda_2
  N_{-\gamma, \beta-\delta}\right) \sin\left( \lambda_1
  N_{-\delta,\beta}\right) E_\alpha ,\label{eq:btrans}
\end{eqnarray} 
where the $N_{\rho_1, \rho_2}$ are the E$_6$ structure constants
\begin{equation}
\left[ E_{\rho_1}, E_{\rho_2}\right] = N_{\rho_1 , \rho_2} E_{\rho_1 +
  \rho_2} ,
\end{equation}
and we have used the facts that $N_{\rho_1, \rho_2}$ is non vanishing
only if $\rho_1 + \rho_2$ is a root and 
\begin{equation}\label{eq:sumnull}
\alpha + \gamma + \delta - \beta = 0 .
\end{equation}
There are several consistency conditions which the structure constants
have to satisfy. These can be found in \cite{carter}. We list them as
they appear for simply laced groups in which case the non vanishing
structure constants can be chosen as $\pm 1$ ($\Phi$ denotes the set of all
root vectors):
\newcounter{prep}
\begin{list}{\bf
    (\roman{prep})}{\usecounter{prep}\rightmargin1.5cm}\itemsep=0pt
\item $N_{\rho,\sigma} = - N_{\sigma, \rho}$ for $\rho,\sigma \in
  \Phi$
\item $N_{\rho_1,\rho_2} = N_{\rho_2,\rho_3} = N_{\rho_3,\rho_1}$ if
  $\rho_1, \rho_2, \rho_3 \in \Phi$ and $\rho_1 + \rho_2 + \rho_3 =
  0$.
\item $N_{\rho, \sigma} = - N_{-\rho,-\sigma}$ for $\rho, \sigma \in
  \Phi$. 
\item $N_{\rho_1,\rho_2} N_{\rho_3,\rho_4} + N_{\rho_2, \rho_3}
  N_{\rho_1,\rho_4} + N_{\rho_3,\rho_1}N_{\rho_2,\rho_4} = 0$ if
  $\rho_1, \rho_2, \rho_3, \rho_4 \in \Phi$ and $\rho_1 + \rho_2 +
  \rho_3 + \rho_4 = 0$ and no pairs are opposite.
\end{list}
These conditions follow essentially from the Jacobi identity (see
\cite{carter}). 
We can use them to find a consistent choice for the constants
appearing in (\ref{eq:atrans}) and (\ref{eq:btrans}). By taking $(\rho_1,
\rho_2, \rho_3)$ as $(\alpha, \gamma, \delta - \beta)$ and $(\alpha+
\delta, \gamma, -\beta)$ one finds from (ii)
\begin{equation}
N_{\gamma, \delta - \beta} = N_{\alpha, \gamma} \,\,\, \mbox{and} \,\,\,
N_{\alpha+ \delta, \gamma} = N_{\gamma, -\beta} .
\end{equation}
Eq.\ (\ref{eq:sumnull}) applied to (iv) yields
\begin{equation}
N_{\alpha,\gamma} N_{\delta, - \beta} + N_{\delta, \alpha}N_{\gamma,
  -\beta} =0,
\end{equation}
which we choose to solve by (using also (iii)) 
\begin{equation}
N_{\alpha,\gamma} = N_{\delta, - \beta} = N_{\delta, \alpha} =
N_{-\gamma,\beta} = 1 .
\end{equation}
After all the structure constants occurring in (\ref{eq:atrans}) and
(\ref{eq:btrans}) have been fixed we find
\begin{eqnarray}
t\left( E_\alpha - E_\beta\right) t^{-1} & = & \left(\cos \lambda_1 \cos
\lambda_2 + \sin \lambda_1 \sin \lambda_2\right)\left( E_\alpha
-E_\beta\right) \nonumber \\
& & + i \left( \sin \lambda_1 \cos \lambda_2 - \sin\lambda_2 \cos
\lambda_1\right) \left( E_{\alpha+\delta} - E_{\alpha + \gamma}\right)
,
\end{eqnarray}
establishing that $E_\alpha - E_\beta$ is indeed invariant if
$\lambda_1 = \lambda_2$. Note also, that the present calculation shows
$E_\alpha + E_\beta$ to be invariant if $\lambda_1 = -\lambda_2$.

\end{appendix}


\begin{thebibliography}{24}
%
\bibitem{Dixon:1985jw}
  L.~J.~Dixon, J.~A.~Harvey, C.~Vafa and E.~Witten,
  Nucl.\ Phys.\ B {\bf 261} (1985) 678.
%
\bibitem{Dixon:1986jc}
  L.~J.~Dixon, J.~A.~Harvey, C.~Vafa and E.~Witten,
  Nucl.\ Phys.\ B {\bf 274} (1986) 285.
%
\bibitem{Ibanez:1986tp}
L.~E.~Ib\'{a}\~{n}ez, H.~P.~Nilles and F.~Quevedo,
Phys.\ Lett.\ B {\bf 187} (1987) 25.
%
\bibitem{Ibanez:1987sn}
  L.~E.~Ib\'{a}\~{n}ez, J.~E.~Kim, H.~P.~Nilles and F.~Quevedo,
  Phys.\ Lett.\ B {\bf 191}, 282 (1987).
%
\bibitem{Bailin:1986pd}
  D.~Bailin, A.~Love and S.~Thomas,
  Phys.\ Lett.\ B {\bf 188} (1987) 193.
%
\bibitem{Bailin:1987pf}
  D.~Bailin, A.~Love and S.~Thomas,
  Nucl.\ Phys.\ B {\bf 288} (1987) 431.
%
\bibitem{Bailin:1987xm}
  D.~Bailin, A.~Love and S.~Thomas,
  Phys.\ Lett.\ B {\bf 194} (1987) 385.
%
\bibitem{Bailin:1987dm}
  D.~Bailin, A.~Love and S.~Thomas,
  Mod.\ Phys.\ Lett.\ A {\bf 3} (1988) 167.
%
\bibitem{Ibanez:1987pj}
L.~E.~Ib\'{a}\~{n}ez, J.~Mas, H.~P.~Nilles and F.~Quevedo,
Nucl.\ Phys.\ B {\bf 301} (1988) 157.
%
\bibitem{Casas:1987us}
  J.~A.~Casas, E.~K.~Katehou and C.~Mu\~{n}oz,
  Nucl.\ Phys.\ B {\bf 317} (1989) 171.
%
\bibitem{Casas:1988se}
  J.~A.~Casas and C.~Mu\~{n}oz,
  Phys.\ Lett.\ B {\bf 209} (1988) 214.
%
\bibitem{Casas:1988hb}
  J.~A.~Casas and C.~Mu\~{n}oz,
  Phys.\ Lett.\ B {\bf 214} (1988) 63.
%
\bibitem{Katsuki:1988ku}
  Y.~Katsuki, Y.~Kawamura, T.~Kobayashi and N.~Ohtsubo,
  Phys.\ Lett.\ B {\bf 212} (1988) 339.
%
\bibitem{Katsuki:1989kd}
  Y.~Katsuki, Y.~Kawamura, T.~Kobayashi, N.~Ohtsubo and K.~Tanioka,
  Prog.\ Theor.\ Phys.\  {\bf 82} (1989) 171.
%
\bibitem{Katsuki:1989ra}
  Y.~Katsuki, Y.~Kawamura, T.~Kobayashi, N.~Ohtsubo, Y.~Ono and K.~Tanioka,
  Phys.\ Lett.\ B {\bf 227} (1989) 381.
%
\bibitem{Hwang:2002hg}
  K.~w.~Hwang and J.~E.~Kim,
  Phys.\ Lett.\ B {\bf 540} (2002) 289
  [arXiv:hep-ph/0205093].
%
\bibitem{Kim:2003ch}
J.~E.~Kim,
Phys.\ Lett.\ B {\bf 564} (2003) 35
[arXiv:hep-th/0301177].
%
\bibitem{Choi:2003pq}
K.~S.~Choi, K.~Hwang and J.~E.~Kim,
Nucl.\ Phys.\ B {\bf 662} (2003) 476
[arXiv:hep-th/0304243].
%
\bibitem{Choi:2003ag}
K.~S.~Choi and J.~E.~Kim,
Phys.\ Lett.\ B {\bf 567} (2003) 87
[arXiv:hep-ph/0305002].
%
\bibitem{Kim:2003hr}
J.~E.~Kim,
JHEP {\bf 0308} (2003) 010
[arXiv:hep-ph/0308064].
%
\bibitem{Kim:2004pe}
J.~E.~Kim,
Phys.\ Lett.\ B {\bf 591} (2004) 119
[arXiv:hep-ph/0403196].
%
\bibitem{Giedt:2001zw}
  J.~Giedt,
  Annals Phys.\  {\bf 297} (2002) 67
  [arXiv:hep-th/0108244].
%
\bibitem{Giedt:2003an}
J.~Giedt,
Nucl.\ Phys.\ B {\bf 671} (2003) 133
[arXiv:hep-th/0301232].
%
\bibitem{Giedt:2005vx}
J.~Giedt, G.~L.~Kane, P.~Langacker and B.~D.~Nelson,
Phys.\ Rev.\ D {\bf 71} (2005) 115013
[arXiv:hep-th/0502032].
%
\bibitem{Choi:2004wn}
K.~S.~Choi, S.~Groot Nibbelink and M.~Trapletti,
JHEP {\bf 0412} (2004) 063
[arXiv:hep-th/0410232].
%
\bibitem{Quevedo:1996sv}
F.~Quevedo,
arXiv:hep-th/9603074.
%
\bibitem{Faraggi:1992fa}
A.~E.~Faraggi,
Nucl.\ Phys.\ B {\bf 387} (1992) 239
[arXiv:hep-th/9208024].
%
\bibitem{Faraggi:1991jr}
A.~E.~Faraggi,
Phys.\ Lett.\ B {\bf 278} (1992) 131.
%
\bibitem{Faraggi:2004rq}
A.~E.~Faraggi, C.~Kounnas, S.~E.~M.~Nooij and J.~Rizos,
Nucl.\ Phys.\ B {\bf 695} (2004) 41
[arXiv:hep-th/0403058].
%
\bibitem{Donagi:2004ht}
R.~Donagi and A.~E.~Faraggi,
Nucl.\ Phys.\ B {\bf 694} (2004) 187
[arXiv:hep-th/0403272].
%
\bibitem{Faraggi:2004xf}
A.~E.~Faraggi,
arXiv:hep-th/0411118.
%
\bibitem{Kawamura:2000ev}
Y.~Kawamura,
Prog.\ Theor.\ Phys.\  {\bf 105} (2001) 999
[arXiv:hep-ph/0012125].
%
\bibitem{Kawamura:2000ir}
Y.~Kawamura,
Prog.\ Theor.\ Phys.\  {\bf 105} (2001) 691
[arXiv:hep-ph/0012352].
%
\bibitem{Altarelli:2001qj}
G.~Altarelli and F.~Feruglio,
Phys.\ Lett.\ B {\bf 511} (2001) 257
[arXiv:hep-ph/0102301].
%
\bibitem{Kawamoto:2001wm}
T.~Kawamoto and Y.~Kawamura,
arXiv:hep-ph/0106163.
%
\bibitem{Hebecker:2001wq}
A.~Hebecker and J.~March-Russell,
Nucl.\ Phys.\ B {\bf 613} (2001) 3
[arXiv:hep-ph/0106166].
%
\bibitem{Asaka:2001eh}
T.~Asaka, W.~Buchm\"uller and L.~Covi,
Phys.\ Lett.\ B {\bf 523} (2001) 199
[arXiv:hep-ph/0108021].
%
\bibitem{Hall:2002ea}
L.~J.~Hall and Y.~Nomura,
Annals Phys.\  {\bf 306} (2003) 132
[arXiv:hep-ph/0212134].
%
\bibitem{Forste:2004ie}
S.~F\"orste, H.~P.~Nilles, P.~K.~S.~Vaudrevange and A.~Wingerter,
Phys.\ Rev.\ D {\bf 70} (2004) 106008
[arXiv:hep-th/0406208].
%
\bibitem{Kobayashi:2004ud}
T.~Kobayashi, S.~Raby and R.~J.~Zhang,
Phys.\ Lett.\ B {\bf 593} (2004) 262
[arXiv:hep-ph/0403065].
%
\bibitem{Kobayashi:2004ya}
T.~Kobayashi, S.~Raby and R.~J.~Zhang,
Nucl.\ Phys.\ B {\bf 704} (2005) 3
[arXiv:hep-ph/0409098].
%
\bibitem{Buchmuller:2004hv}
W.~Buchm\"uller, K.~Hamaguchi, O.~Lebedev and M.~Ratz,
Nucl.\ Phys.\ B {\bf 712} (2005) 139
[arXiv:hep-ph/0412318].
%
\bibitem{Buchmuller:2005jr}
W.~Buchm\"uller, K.~Hamaguchi, O.~Lebedev and M.~Ratz,
arXiv:hep-ph/0511035.
%
\bibitem{Ibanez:1987xa}
  L.~E.~Ib\'{a}\~{n}ez, H.~P.~Nilles and F.~Quevedo,
  Phys.\ Lett.\ B {\bf 192} (1987) 332.
%
\bibitem{Forste:2005rs}
  S.~F\"orste, H.~P.~Nilles and A.~Wingerter,
  Phys.\ Rev.\ D {\bf 72} (2005) 026001
  [arXiv:hep-th/0504117].
%
\bibitem{Font:1988tp}
A.~Font, L.~E.~Ib\'{a}\~{n}ez, H.~P.~Nilles and F.~Quevedo,
Nucl.\ Phys.\ B {\bf 307} (1988) 109
[Erratum-ibid.\ B {\bf 310} (1988) 764].
%
\bibitem{Font:1988mm}
A.~Font, L.~E.~Ib\'{a}\~{n}ez, H.~P.~Nilles and F.~Quevedo,
Phys.\ Lett.\  {\bf 210B} (1988) 101
[Erratum-ibid.\ B {\bf 213} (1988) 564].
%
\bibitem{Gmeiner:2002es}
F.~Gmeiner, S.~Groot Nibbelink, H.~P.~Nilles, M.~Olechowski and
M.~G.~A.~Walter, 
Nucl.\ Phys.\ B {\bf 648} (2003) 35
[arXiv:hep-th/0208146].
%
\bibitem{Hebecker:2003jt}
A.~Hebecker and M.~Ratz,
Nucl.\ Phys.\ B {\bf 670} (2003) 3
[arXiv:hep-ph/0306049].
%
\bibitem{Akin} A.\ Wingerter, PhD Thesis, Bonn University, June 2005, http://hss.ulb.uni-bonn.de:90/ulb\_bonn/diss\_online/math\_nat\_fak/2005/wingerter\_akin/index.htm  
%
\bibitem{Nilles:2004ej}
H.~P.~Nilles,
arXiv:hep-th/0410160.
%
\bibitem{Schellekens:1987ij}
  A.~N.~Schellekens and N.~P.~Warner,
  Nucl.\ Phys.\ B {\bf 308} (1988) 397.
%
\bibitem{Bouwknegt:1988hn}
  P.~Bouwknegt,
  J.\ Math.\ Phys.\  {\bf 30} (1989) 571.
%
\bibitem{carter}
R.\ W.\ Carter, {\it ``Simple Groups of Lie Type,''} John Wiley \&
Sons (1972) 331 p. (Pure and Applied Mathematics 28).
%
\bibitem{Zee:2003mt}
  A.~Zee,
 {\it  ``Quantum field theory in a nutshell,''} Princeton
  Univ. Pr. (2003) 518 p.  
\end{thebibliography}
\end{document}